\documentstyle[11pt,epsf]{article}
\newcommand{\beq}{\begin{equation}}
\newcommand{\eeq}{\end{equation}}
\newcommand{\eq}[1]{(\ref{#1})}
\newcommand{\bea}{\begin{eqnarray}}
\newcommand{\eea}{\begin{eqnarray}}
\newcommand{\beas}{\begin{eqnarray*}}
\newcommand{\eeas}{\begin{eqnarray*}}
\newcommand{\postscript}[2]
   {\setlength{\epsfxsize}{#2\hsize}
   \centerline{\epsfbox{#1}}}

\makeatletter\@addtoreset {equation}{section}\makeatother
\title{A scalar nonlocal bifurcation of solitary waves for coupled nonlinear 
Schr\"odinger systems}
\author{Alan Champneys\thanks{Department of Engineering Mathematics,
University of Bristol, Bristol BS8 1TR {\tt a.r.champneys@bris.ac.uk}} \,
and Jianke Yang\thanks{Department of Mathematics and Statistics, 
The University of Vermont, 16 Colchester Avenue, Burlington, VT 05401, USA
{\tt jyang@emba.uvm.edu} } }
\date{ }
\begin{document}
\maketitle
\begin{abstract}

An explanation is given for previous numerical results which suggest a
certain bifurcation of `vector solitons' from scalar
(single-component) solitary waves in coupled nonlinear Schr\"{o}dinger
(NLS) systems.  The bifurcation in question is nonlocal in the sense
that the vector soliton does not have a small-amplitude component, but
instead approaches a solitary wave of one component with two
infinitely far-separated waves in the other component.  Yet, it is
argued that this highly nonlocal event can be predicted from a purely
local analysis of the central solitary wave alone. Specifically the
linearisation around the central wave should contain asymptotics which
grow at precisely the speed of the other-component solitary waves on
the two wings.  This approximate argument is supported by both a
detailed analysis based on matched asymptotic expansions, and
numerical experiments on two example systems.  The first is the usual
coupled NLS system involving an arbitrary ratio between the self-phase
and cross-phase modulation terms,
and the second is a coupled NLS system with saturable nonlinearity
that has recently been demonstrated to support stable multi-peaked
solitary waves. The asymptotic analysis further reveals that 
when the curves which define the proposed criterion for scalar
nonlocal bifurcations intersect with boundaries of certain
local bifurcations, the nonlocal bifurcation could turn from 
scalar to non-scalar at the intersection. This phenomenon is 
observed in the first example.  Lastly, we have also selectively tested
the linear stability of several solitary waves just born out of 
scalar nonlocal bifurcations. We found that they are linearly unstable. 
However, they can lead to stable solitary waves through parameter continuation. 

\end{abstract}



\section{Introduction}
Solitary waves play an important role in the solution dynamics of
nonlinear evolution equations. If the solitary waves are stable, 
they often emerge as final states in an initial-value problem. 
Even if these waves are unstable, the mere existence of such waves
has important implications for solution evolutions. 
In recent years, it has been discovered that 
complicated solitary waves could bifurcate from simple solitary waves. 

{\em Local bifurcations} of `wave and daughther waves'
(`vector solitons') from single-component waves (`scalar
solitons') have been studied in various forms of coupled nonlinear
Schr\"odinger (NLS) systems \cite{Eleonsky,HaSh:94,KoBuStChSa:00,
OsKi:99,Ya:97,Be:96}.
The bifurcation is {\em local} in that
bifurcated solitary wave is infinitesimally close to the original
solitary wave (as a graph) at the point of bifurcation.  The condition
for such a local bifurcation to occur is based on the linearisation
around the single-component pulse having a solution with purely
decaying asymptotics at infinity (see Secs.~\ref{sec:1} and
\ref{sec:ex1} below).

{\em Nonlocal bifurcations} are where the bifurcated solitary wave is
not infinitessimally close to the original wave at the point of
bifurcation.  These have been reported numerically in the usual
coupled NLS equations (the generalised Manakov system, involving an
arbitrary ratio between the self-phase and cross-phase studied in
Sec.~\ref{sec:ex1} below, henceforth referred to as the CNLS system)
\cite{HaSh:94,Ya:97,Ya:98}, and in second- and third-harmonic
generation systems \cite{KoBuStChSa:00,Ye:01,YeChMc:99}.  If the
bifurcated wave looks like several vector solitons glued together,
that nonlocal bifurcation has also been treated analytically by an
asymptotic 
tail-matching method \cite{Ya:98} (see Sec.~\ref{sec:NLS_analysis} for
application of this method in the context of this paper).  General
mechanisms have also been identified that lead to such vector
non-local bifurcations in classes of coupled NLS systems, such as
eigenvalue degeneracy or the existence of a local bifurcation
\cite{Ya:98,Ye:01,YeSaJo:00}.  The linear stability of these multiple-pulsed
vector solitons for the CNLS system has been studied in \cite{Ya:01},
and it has been shown that such states are linearly unstable.
However, if the bifurcated wave is glued together by scalar (i.e.,
single-component) solitary waves, no analysis has been performed to
our best knowledge. In this paper, we focus on this {\em scalar}
nonlocal bifurcation. We will show that it is closely related to local
bifurcations, and can be treated on the same footing.

The criterion we propose for a scalar nonlocal bifurcation (in
Sec.~\ref{sec:1}) is that the solution of the linearized equation
around the central single-component pulse should have only purely growing
asymptotics instead of purely decaying asymptotics at infinity. In the
following Sections, we test this criterion against two example systems.  The
first, in Sections \ref{sec:ex1}, is the CNLS system.  The second, in
Section \ref{sec:ex2}, is a coupled NLS system with saturable
nonlinearity.  In both cases, good agreement is obtained between our
bifurcation condition and the numerics.  Near the bifurcation point,
we have also developed a detailed asymptotic analysis based on the
above-mentioned tail-matching method, which is performed for the CNLS
system in Sec.~\ref{sec:NLS_analysis}.  This analysis produces an
explicit formula for the spacing between scalar solitons being pieced
together, and this formula agrees well with the numerics.  An
extension of that analytical theory to more general coupled NLS
systems, such as our second example studied in Sec.~\ref{sec:ex2} is
straightforward, and will be omitted in this paper.  Lastly, we
discover in the course of our asymptotic analysis that if the
curve which defines our proposed criterion for scalar nonlocal
bifurcations intersects with boundaries of certain local bifurcations, 
the nonlocal bifurcation may turn from scalar to non-scalar at the
intersection. This phenomenon indeed occurs in the CNLS system. 
We have also studied the linear stability of solitary waves just born out of
scalar nonlocal bifurcations. The results suggest that these waves are always
linearly unstable. However, they can lead to stable solitary waves 
through parameter continuation in the second model. 

\section{A geometric argument}
\label{sec:1}
Consider a general system of ordinary differential equations (ODEs) of the form
\begin{equation}
\label{genu}
u_{xx}-u+f_1(u,v)u=0, 
\end{equation}
\begin{equation} \label{genv}
v_{xx}-\omega^2 v+f_2(u,v)v=0, 
\end{equation}
where $0<\omega<1$.
It is assumed that $f_1$ and $f_2$ are smooth nonlinear functions of
their arguments which vanish as $(u,v)\to (0,0)$ and 
may well depend on other system parameters.
Moreover they are such that the problem with $v\equiv 0$ or with
$v\equiv 0$, which are both invariant subspaces within the four-dimensional
phase space of the ODE system (\ref{genu}),(\ref{genv}), contain
even homoclinic solutions
\beq
(u(x),v(x))=(u_h(x),0), \quad   u_h(-x)=u_h(x), \qquad u\rightarrow 0 \quad 
\mbox{as} \quad x \to \pm \infty,
\label{genupulse}
\eeq
\beq
(u(x),v(x))=(0,v_h(x)), \quad   v_h(-x)=v_h(x), \qquad v\rightarrow 0 \quad
\mbox{as} \quad x \to \pm \infty.
\label{genvpulse}
\eeq
In what follows we shall use the terms `pulse'
`homoclinic to the origin', `solitary wave' and `soliton' entirely
synonymously.
We shall refer to the invariant subspace homoclinic 
solutions \eq{genupulse} and \eq{genvpulse} 
as {\em scalar} $u$- and $v$-pulses respectively. These
scalar solitons are contrasted with {\em vector solitons} which are
homoclinic solutions that have non-zero $u$ and $v$ components. 
Note that all homoclinic solutions to the origin are generic, 
that is they persist under parameter perturbation, 
since the system is both reversible and Hamiltonian
\cite{De:76b}. 

Now consider the linearisation of (\ref{genu}),(\ref{genv})
around the $u$-pulse 
\begin{equation}
\label{genlinu}
u_{xx}=u+ \left\{\frac{\partial }{\partial u}
f_1[u_h(x),0] u_h(x)+f_1[u_h(x),0]\right\}u, \quad 
\end{equation}
\begin{equation}
\label{genlinv}
v_{xx}=\omega^2 v +  f_2[u_h(x),0)]v.
\end{equation}
Note that the linear equations decouple. So  let us 
look at the specific class of solutions to this linear problem
which have $u=0$. Now we have to simply solve the second equation
(\ref{genlinv}). The general asymptotics of such solutions satisfy
\begin{equation}
\label{gentail}
v \to c_1^{\pm} e^{-\omega x} + c_2^{\pm} e^{\omega x} + o\left (e^{-\omega |x|}\right ) \:
\mbox{as} \: |x| \to \pm \infty
\end{equation}
for some constants $c_1^\pm$ and $c_2^\pm$. Note that we are able to
write the $o(\cdot)$-term by the assumption that $0<\omega<1$ so that
the asymptotics of $f_2(u(x),0)$ decay more rapidly than
$\exp(-\omega |x|)$.

\begin{figure}[p]
\begin{center}
\parbox{16cm}{\postscript{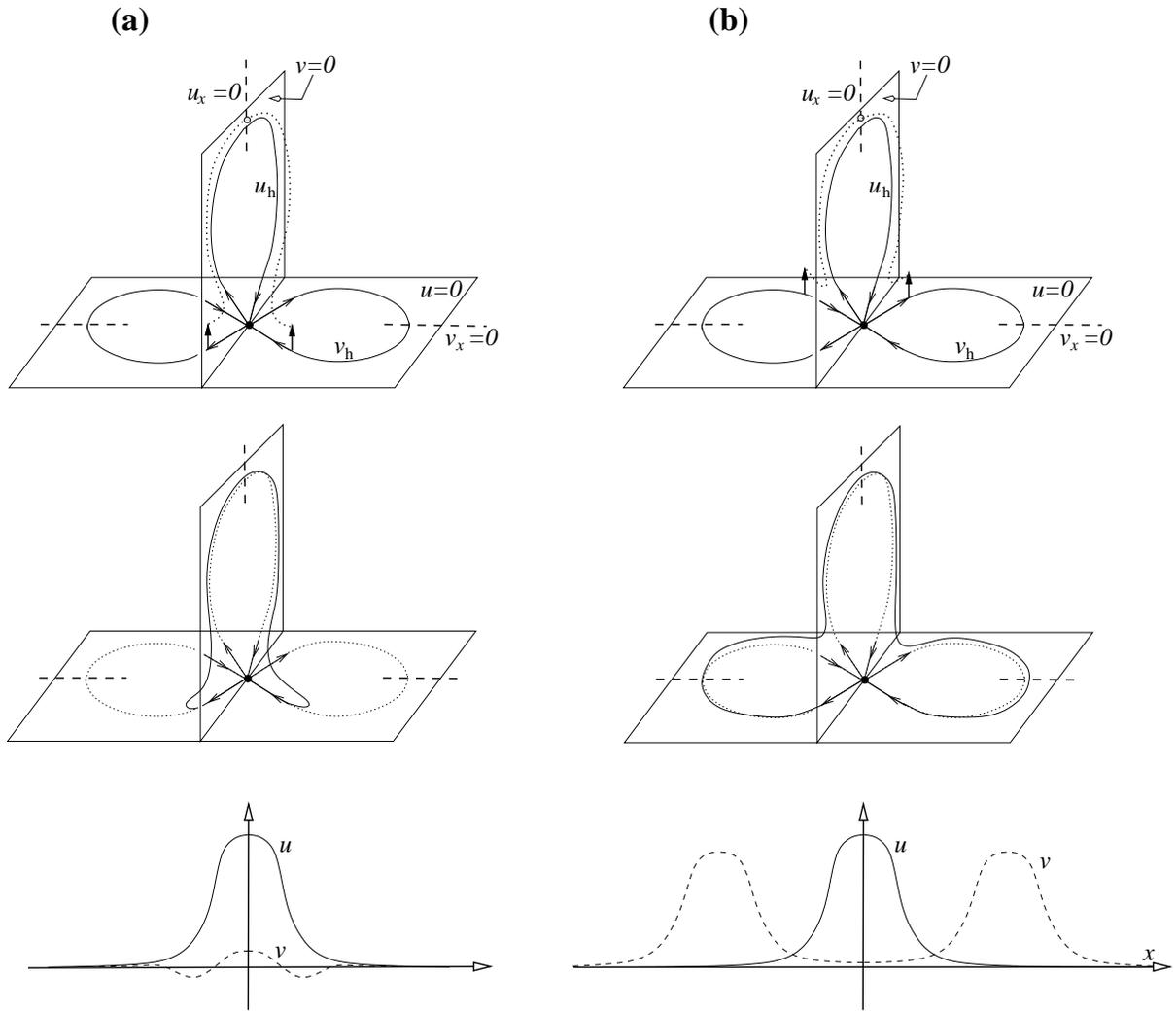}{1.0}}
\caption{Sketch figure defining (a) scalar nonlocal and (b) local bifurcations. In each
case the top panel depicts schematically the invariant planes $\{u\equiv 0\}$
and $\{v \equiv 0\}$ and the solution of the linearised problem around $u_h$.
The lower two panels depict the corresponding bifurcated vector solitons
in phase space and as graphs.}
\label{Fig:sketch}
\end{center}
\end{figure}

Consider even solutions \eq{gentail} of \eq{genlinv},
hence 
$c_i^+=c_i^-:=c_i$, $i=1,2$. This 
defines a unique solution (up to scale) for the linear 
initial-value problem (\ref{genlinv})
for $x\geq 0$. Now suppose that at a particular
value of $\omega$, this solution had a particular tail asymptotics
(\ref{gentail}) with $c_2=0$ (see Fig.~\ref{Fig:sketch}(b)). Then
the solution to the linear problem would be localized. Going back to
the fully nonlinear problem, by standard bifurcation theory results,
we have satisfied the necessary condition for the {\em local}
bifurcation of a wave and daughter wave consisting of the mother
$u$-pulse and a small-amplitude $v$-component.
The vanishing of $c_2$ is a codimension-one
condition, and hence local bifurcations will lie on lines in a
parameter plane.  As already mentioned in the Introduction, the 
existence of such bifurcations in coupled NLS
systems have been established by a number of authors.

Suppose instead that we find a solution at some parameter value that
satisfies $c_1=0$. Note that this describes a pure exponentially
growing solution to leading order as $x \to \pm \infty$.  
See Fig.~\ref{Fig:sketch}(a). However, consider the nonlinear implications of
this within the four-dimensional phase space $\{(u,u',v,v'): u=u'=0\}$
of the ODE \eq{genu},\eq{genv}.  We have found an initial condition
for an even solution that is an infinitesimal perturbation of
$u_h(0)$, and which is attracted as $x \to \infty$ towards the
invariant plane $V:=\{(u,u',v,v'): u=u'=0\}$. Moreover, this rate of
attraction is faster ($ \sim \exp( x)$) than the exponential
contraction or expansion with that plane ($ \sim \exp ( \pm \omega x
)$) near the origin.  Hence the condition that $0<\omega<1$ ensures
that the invariant plane $V$ is normally hyperbolic..  Using standard
results for normally hyperbolic manifolds, the eventual behaviour of
this perturbed trajectory is governed by its behaviour on $V$. The
fact $c_1=0$ implies that the trajectory is attracted onto
the local unstable manifold within $V$.  But the local unstable
manifold on $V$ is precisely the piece of trajectory that forms the
$v$-pulse solution $v_h$. Also since we are talking about an
infinitesimal perturbation to the underlying pulse $u_h$, the time
taken to be attracted to $v_h$ in this way is arbitrarily long. Hence
we have the scalar nonlocal bifurcation of two $v_h$ pulses `at $\pm
\infty$' as depicted in Fig.~\ref{Fig:sketch}(a).

The above is only a plausible argument, but it is highly appealing
from an intuitive point of view. In Sec.~\ref{sec:NLS_analysis}, 
we will develop an in-depth asymptotic theory for this scalar nonlocal bifurcation
in the CNLS equations. The results of that theory fully
support the above intuitive argument. 

Before proceeding to the examples, let us make a few short remarks here. 
\begin{enumerate}
\item
First, note the need to assume that
$0<\omega<1$. This was required in order to make the $u=0$ invariant
plane normally hyperbolic, or equivalently to assume that the asymptotic
attraction onto this plane was $o(e^{-\omega x})$. If this condition is violated,
then there is no sense in which the perturbed trajectory converges only to the 
unstable manifold within $V$ and hence the argument fails.
\item
Second, it is
interesting to note that our analysis suggest that local and scalar nonlocal
bifurcations can be treated on an equal footing. One requires that
$c_1$ vanishes, the other that $c_2$ vanishes. It is perfectly possible
to imagine a scenario where, as a parameter is varied, the even solution
$v(x)$ to the linear problem (\ref{genlinv}) generates extra internal
oscillations. If it does so in a smooth way, then it is clear 
that it must pass repeatedly through successive zeros of 
$c_1$ and $c_2$. Hence one would find each  scalar nonlocal bifurcation
sandwiched between two successive local bifurcations. 

\item In fact, in the above it is not quite enough to assume
that $c_1$ vanishes for a scalar nonlocal bifurcation. We must have that
$c_2$ has the correct sign to attach to the component of
the unstable manifold of the nonlinear equation that contains
the pure $v$-pulse. In all the examples below, the pure-$v$
equation is odd and hence both $v_h$ and $-v_h$ are solutions.
Thus either sign of $c_2$ will lead to a scalar nonlocal bifurcation.
\item Finally, in systems with odd nonlinearity, 
the above arguments can be repeated 
to find anti-symmetric scalar nonlocal bifurcations
where the two daughter waves at infinity are $v_h$ and $-v_h$. 
\end{enumerate}
 
We now turn to two examples to test the validity of this approximate
reasoning.

\section{Example 1: the CNLS equations}
\label{sec:ex1}

The usual coupled nonlinear Schr\"{o}dinger (CNLS) equations
may be written in dimensionless form as 
\begin{eqnarray}
iU_t + U_{xx} + (|U|^2 + \beta |V|^2) U & = & 0, \label{CNLS_PDE1} \\
iV_t + V_{xx} + (|V|^2 + \beta |U|^2) V & = & 0. \label{CNLS_PDE2}
\end{eqnarray}
They have been used to describe the interaction between wave packets
in dispersive conservative media, and also the
interaction between orthogonally polarized components in nonlinear
optical fibres; see \cite{Ya:97,HaSh:94} and references therein. 
Looking for steady solutions of the form
$$ 
U= e^{i \omega_1^2 t} u(x), \quad V= e^{i \omega_2^2 t} v(x),
$$
and performing scaling so that $\omega_1=1$ and
$\omega_2=\omega$,
we arrive at the following set of ODEs 
\begin{equation} \label{u}
u_{xx}-u+(u^2+\beta v^2)u=0, 
\end{equation}
\begin{equation} \label{v}
v_{xx}-\omega^2 v+(v^2+\beta u^2)v=0. 
\end{equation}
Here $\beta$ is a real and positive cross-phase-modulational
coefficient, and $\omega \;(\ge 0)$ is a propagation constant parameter. 
It is noted that if $[u(x; \omega), v(x; \omega]$ is a solution, 
then another solution at propagation constant $\frac{1}{\omega}$
can be obtained via the transformation \cite{Ya:97}
\beq
[\frac{1}{\omega} \hspace{0.05cm} v(\frac{x}{\omega}; \hspace{0.05cm} \omega),\;
\frac{1}{\omega}\hspace{0.05cm} u(\frac{x}{\omega}; \hspace{0.05cm} \omega)]. 
\label{omegatrans}
\eeq
Thus, in this paper, we restrict $0\le \omega \le 1$. 

We note that when $\beta=1$, the partial differential equations (PDEs)
\eq{CNLS_PDE1}, \eq{CNLS_PDE2} 
are called the Manakov system which is integrable \cite{Ma:73}. 
In this case, all solitary waves of the ODE system \eq{u} and \eq{v}
have closed-form analytical expressions \cite{Ak:95}. 
When $\beta=0$, the PDEs are two copies of the single NLS 
equation which is also integrable \cite{Za:71}. 
The solitary waves for $\beta=0$ are simply sech pulses. 
When $\beta\ne 0$ or 1, the structure of solitary waves in this system
is much more complicated. This structure was partially unraveled in
\cite{Eleonsky,AkAn:93,Ak:95,HaSh:94,Ya:97,Ya:98}. 
It is known that local bifurcations occur along curves in
the $(\beta,\omega)$-plane that are given by closed form expressions
(see below). 
These local bifurcations are where wave and daughter-wave
structures are born. 
In other words, at local bifurcations, a small and localized
$v$ component develops from a pure $u$-pulse. 
It was also observed numerically that scalar nonlocal bifurcations
such as described in this paper occur.
That is, passing through the bifurcation is a
single component pulse, for which $v$ is zero everywhere, and
$u$ is given by a sech function. Bifurcating from this is a solution for
which the $u$ component remains about the same, but the $v$
component suddenly develops two pulses which are far 
separated from the central $u$ pulse. These $v$ pulses can be 
symmetrically or antisymmetrically distributed. 
Their sizes jump from zero to a certain
finite size across the bifurcation. 
In this section, we analytically determine the boundaries of these
scalar nonlocal bifurcations through the criterion developed in Sec. 2
and compare them with direct numerical results. 

\subsection{Local and scalar nonlocal bifurcations}

First, we recall the results for local bifurcations in this system
\cite{Eleonsky,Be:96, Ya:97},
assuming that a small $v$-component bifurcates from a pure $u$-pulse.
Thus, at the bifurcation point, the $v$ component is infinitesimally
small. Thus the $u$-component is simply governed by the 
equation 
\begin{equation}
u_{xx}-u+u^3=0, 
\end{equation}
whose homoclinic solution is 
\begin{equation}
u(x)=\sqrt{2}\mbox{sech}\, x. 
\label{upulse}
\end{equation}
According to standard results, a necessary condition of a local bifurcation of
a homoclinic solution with a small-amplitude $v$ component from the 
$u$-pulse
\eq{upulse} is that there is a non-trivial localised solution to the 
linearised problem of the $v$-component. This takes the form of a 
linear Schr\"odinger equation
\begin{equation} \label{schroedinger}
v_{xx}-\omega^2 v+2\beta \mbox{sech}^2 x v=0, 
\end{equation}
and for local bifurcation we require
$$
v\to 0 \quad \mbox{as} \quad |x| \to \pm \infty.
$$
This equation can be solved exactly 
\cite{LaLi:77} as we now explain.
With the variable transformation
\begin{equation}
v=\mbox{sech}^s x \: \psi, \hspace{0.5cm} \xi=\sinh^2x, 
\end{equation}
where 
\begin{equation}\label{sbeta}
s=\frac{\sqrt{1+8\beta}-1}{2}, 
\end{equation}
the Schr\"odinger equation (\ref{schroedinger}) becomes
\begin{equation}
\xi(1+\xi)\phi_{xx}+[(1-s)\xi+\frac{1}{2}]\phi_x +\frac{1}{4}
(s^2-\omega^2)\phi=0, 
\end{equation}
which is a hyper-geometric equation. Its even and odd solutions
are
\begin{equation}
\phi_1=F(\frac{1}{2}\omega-\frac{1}{2}s, 
	 -\frac{1}{2}\omega-\frac{1}{2}s, 
	\frac{1}{2}, -\xi), 
\end{equation}
\begin{equation}
\phi_2=\sqrt{\xi}F(\frac{1}{2}\omega-\frac{1}{2}s+\frac{1}{2}, 
	 -\frac{1}{2}\omega-\frac{1}{2}s+\frac{1}{2}, \frac{3}{2}, -\xi),
\end{equation}
where $F$ is the hyper-geometric function. 
In order for the solution $v_1=\mbox{sech}^s x \phi_1$ 
to decay to zero as $x$ goes to infinity, 
one must have
\begin{equation} \label{cond1}
\frac{1}{2}\omega-\frac{1}{2}s = -n_1, 
\end{equation}
where $n_1$ is a non-negative integer, and 
$n_1<\frac{1}{2}s$. 
Then the $v_1$ solution is
\begin{equation} \label{v1}
v_1=\mbox{sech}^s x \sum_{k=0}^{n_1}
\frac{(-n_1)_k(-\frac{1}{2}\omega-\frac{1}{2}s)_k (-\xi)^k}
			  {(\frac{1}{2})_k k!}, 
\end{equation}
which decays to zero as $\xi^{-\omega/2}$ (i.e., $e^{-\omega x}$). 
In the solution (\ref{v1}), $(a)_k$ is defined as
\begin{equation}
(a)_k\equiv \left\{ \begin{array}{ll} a(a+1)(a+2)\dots (a+k-1), & k>0 \\
		1, & k=0. \end{array}\right.
\end{equation}

In order for solution $v_2=\mbox{sech}^s x \phi_2$ 
to decay to zero as $x$ goes to infinity, one must have 
\begin{equation} \label{cond2}
\frac{1}{2}\omega-\frac{1}{2}s+\frac{1}{2} = -n_2, 
\end{equation}
where $n_2$ is a non-negative integer, and 
$n_2 < \frac{1}{2}(s-1)$. Then the $v_2$ solution is 
\begin{equation} \label{v2}
v_2=\mbox{sech}^s x \sinh x \sum_{k=0}^{n_2} 
\frac{(-n_2)_k(-\frac{1}{2}\omega-
\frac{1}{2}s+\frac{1}{2})_k (-\xi)^k}{(\frac{3}{2})_k k!}, 
\end{equation}
which also decays to zero as $\xi^{-\omega/2}$ (i.e., $e^{-\omega x}$). 

\begin{figure}[p]
\begin{center}
\parbox{12cm}{\postscript{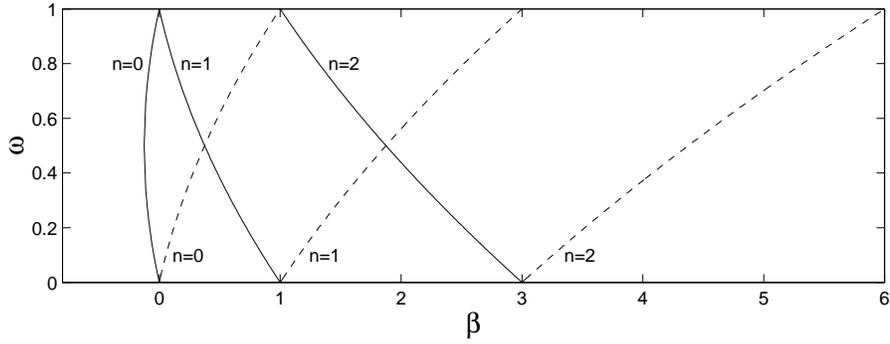}{1.0}}
\caption{Local and scalar nonlocal bifurcation boundaries in the coupled NLS
system (\ref{u}) and (\ref{v}). Scalar nonlocal bifurcation boundaries are
solid lines, and are given by Eq. (\ref{firstcurve}) [with $n=0$] and 
(\ref{nonlocal}) [with $n=1, 2$].  Local bifurcation
boundaries are dashed lines, and are given by Eq.~(\ref{local}). }
\label{Fig:fig1}
\end{center}
\end{figure}

When conditions (\ref{cond1}) and (\ref{cond2}) are combined, 
we find that the boundaries for local bifurcations are
\begin{equation} \label{local}
\omega=\omega^{LB}_n(\beta)=s-n, 
\end{equation}
where $s$ is given by \eq{sbeta},
$n$ is a non-negative integer, and $n<s$. 
The first boundary $(n=0)$ exists for any $\beta \ge 0$; 
the second boundary $(n=1)$ exists only for $\beta \ge 1$; 
the third boundary $(n=2)$ exists only for $\beta \ge 3$; etc. 
The first three boundaries $\omega^{LB}_{0,1,2}$ are plotted in 
Fig.~\ref{Fig:fig1} as dashed
lines for illustration. Note from the above construction that even $n$
corresponds to the existence of symmetric bifurcating waves (even in
both $u$ and $v$) whereas odd $n$ corresponds to anti-symmetric
bifurcation (even in $u$, odd in $v$).

Now, how can we define the scalar nonlocal bifurcations?  It can be 
noted that on the above {\em local} bifurcation boundaries, the
appropriate solution $v_1$ or $v_2$ has the following asymptotic
behavior:
\begin{equation} \label{asym1}
v(x) \longrightarrow \mbox{sgn}^n(x)
\left (c_{n} e^{\omega |x|} +d_n e^{-\omega |x|} +o(e^{\omega |x|})
\right ),   \hspace{0.5cm} \mbox{as} \quad |x| \rightarrow \infty, 
\end{equation}
where $c_n=0$, and $d_n$ is a non-zero constant. 
In other words, this $v$ solution is localized. 
Condition $c_n=0$ in the asymptotics 
(\ref{asym1}) is the condition for local bifurcations in this problem. 

Now, following the arguments laid out in Sec.~\ref{sec:1}, a {\em
scalar nonlocal} bifurcation occurs when the $v$-component of the linearised
equation around the $u$-pulse satisfies conditions at infinity that it
has a purely growing component.  That is, scalar nonlocal bifurcations occur
when one of the $v_1$ and $v_2$ solutions of the Schr\"odinger
equation (\ref{schroedinger}) has the following asymptotic behavior:
\begin{equation} \label{asym2}
v(x) \longrightarrow \mbox{sgn}^j(x) \left 
(\alpha e^{\omega |x|}+\gamma e^{-\omega |x|} 
+o(e^{-\omega |x|}) \right)
\hspace{0.5cm} \mbox{as} \quad |x| \rightarrow \infty,
\end{equation}
with $\gamma=0$ but some non-zero $\alpha$, and $j=0$ or 1 is an
integer indicating the symmetry of the $v$ solution.
Now, a remarkable thing happens. Because it is easy to see  
using the solutions of Eq.\ (\ref{schroedinger}) obtained above, that
condition $\gamma=0$ in the asymptotic equation (\ref{asym2}) is
exactly satisfied on the boundary curves
\begin{equation} \label{nonlocal}
\omega= \omega^{NLB}_n(\beta):=n-s, \qquad 0<n-s<1, 
\end{equation}
where $n$ is a non-negative integer and $s(\beta)$ was defined 
in Eq. (\ref{sbeta}). 
In fact, on these boundaries, 
function $v$ has an unbounded solution 
\begin{equation} \label{v1n}
\hat{v}_1=\mbox{sech}^s x \sum_{k=0}^{n_1}
\frac{(-n_1)_k(\frac{1}{2}\omega-\frac{1}{2}s)_k (-\xi)^k}
                          {(\frac{1}{2})_k k!}
\end{equation}
when $n=2n_1$ is even, and an unbounded solution
\begin{equation} \label{v2n}
\hat{v}_2=\mbox{sech}^s x \sinh x \sum_{k=0}^{n_2} 
\frac{(-n_2)_k(\frac{1}{2}\omega-
\frac{1}{2}s+\frac{1}{2})_k (-\xi)^k}{(\frac{3}{2})_k k!}, 
\end{equation}
when $n=2n_2+1$ is odd. 
The asymptotic behaviors of these solutions are
\begin{equation} \label{vasym}
\hat{v}(x) \longrightarrow \mbox{sgn}^n(x) \left (g_n e^{\omega |x|}+
h_n e^{(\omega-2)|x|} \right ), \hspace{0.5cm} |x| \rightarrow \infty, 
\label{NLB_cond}
\end{equation}
where 
\begin{equation}\label{gn}
g_n=\left\{\begin{array}{ll}
\frac{2^s \left(\frac{1}{2}\omega-\frac{1}{2}s\right)_{n_1}}{4^{n_1} 
\left(\frac{1}{2}\right)_{n_1}}, & n=2n_1, \\
& \\
\frac{2^{s-1} \left(\frac{1}{2}\omega-\frac{1}{2}s+\frac{1}{2}\right)_{n_2}}{4^{n_2} 
\left(\frac{3}{2}\right)_{n_2}}, & n=2n_2+1, \end{array}\right.
\end{equation}
and $h_n$ is another constant which can be easily calculated. 
When $\omega<1$, the second term in \eq{NLB_cond} decays faster than 
$e^{-\omega |x|}$. Thus the coefficient $\gamma$ in the asymptotics
(\ref{asym2}) is zero on these boundaries. 

The boundary curve for scalar nonlocal bifurcations (\ref{nonlocal}) 
can be written alternatively  as 
\begin{equation} \label{nonlocal2}
\beta=\beta_n^{NLB}(\omega)=\frac{1}{8}\left\{ (2n-2\omega+1)^2-1 \right\}. 
\end{equation}
These  boundaries are plotted in Fig.~\ref{Fig:fig1} 
as solid
lines for comparison with boundaries of local bifurcations (dashed lines), 
which are given according to \eq{local} by
\begin{equation} \label{local2}
\beta=\beta_n^{LB}(\omega)=\frac{1}{8}\left\{ (2n+2\omega+1)^2-1 \right\}. 
\end{equation}
Hence, by construction, the 
curves of nonlocal bifurcations simply represent the continuation of 
local bifurcation curves 
through $\omega=0$ (and mapped back up via $\omega \to -\omega$, since
only $\omega^2$ appears in the equations).
In particular, at the singular value $\omega=0$ we have
that $\omega^{NLB}_n=\omega^{LB}_n$.

Lastly, we note that the first scalar nonlocal bifurcation curve
on the left of Fig.~\ref{Fig:fig1} needs a little special treatment. 
In fact, this solid curve is given by equation
\begin{equation}\label{firstcurve}
\omega=\left\{ \begin{array}{ll} -s,  & -\frac{1}{8}\le \beta\le 0\; \mbox{and}\; 
\omega\le \frac{1}{2}, \\
\frac{1}{2}\left[1+\sqrt{1+8\beta}\right], & -\frac{1}{8}\le \beta\le 0\; \mbox{and}\;
\omega\ge \frac{1}{2}. \end{array} \right.
\end{equation}
In other words, the lower branch of this curve is as given by Eq. (\ref{nonlocal}) with
$n=0$, but its upper branch is given by a different function. 
It can be shown that on this upper branch, 
the solution of the Schr\"odinger equation (\ref{schroedinger})
also has the asymptotics (\ref{asym2}) with $\gamma=j=0$. 
This curve is the only scalar nonlocal bifurcation boundary 
whose functional form is partially different from (\ref{nonlocal}).

\subsection{Numerical results}

So we have found curves on which our proposed condition for scalar nonlocal
bifurcations derived in Sec.~\ref{sec:1} is satisfied. It remains
to be seen what happens to the fully nonlinear equations for this
example along such curves. In Sec.~\ref{sec:NLS_analysis} below
we shall consider this problem via asymptotic analysis. In this section
we turn to numerical methods. 

\begin{figure}[p]
\begin{center}
\parbox{14cm}{\postscript{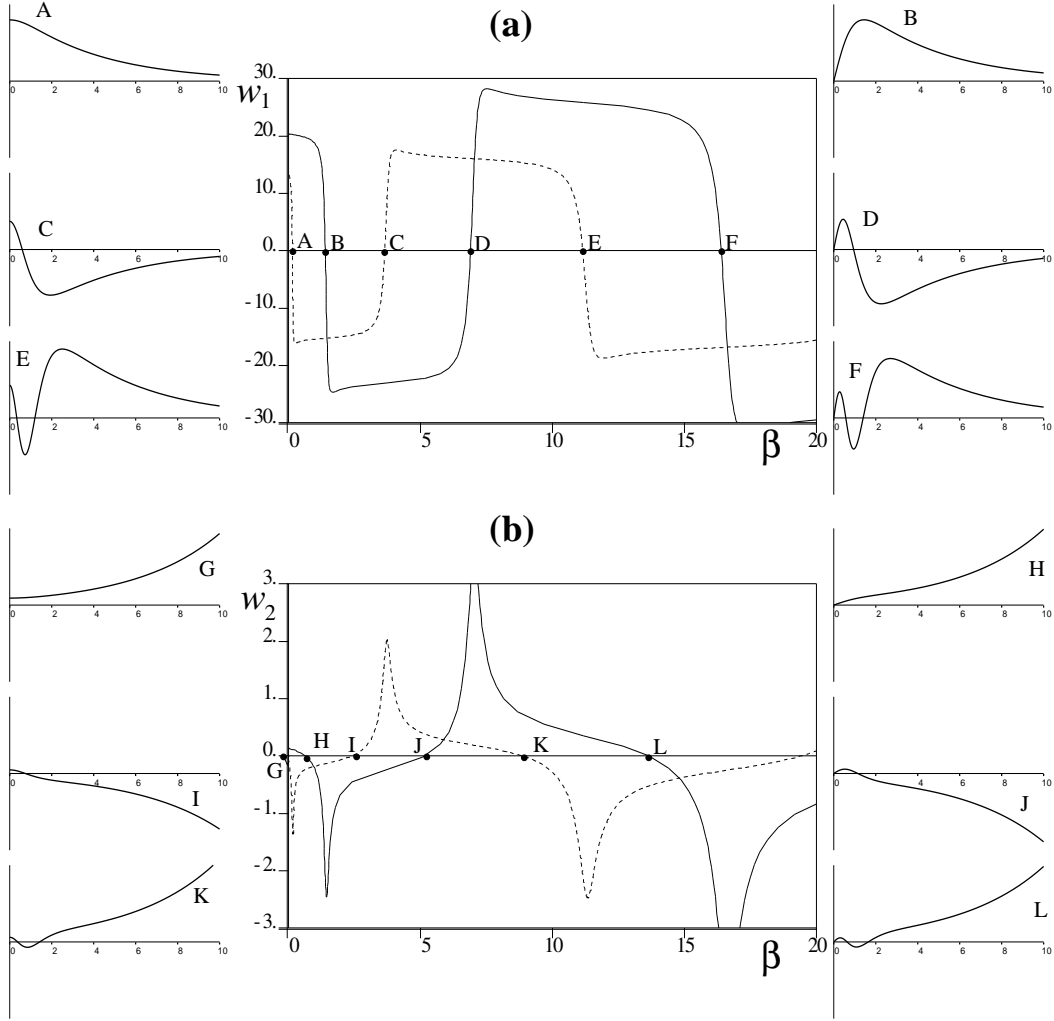}{1.0}}
\caption{Curves of (a) $w_1(\beta)$ and (b) $w_2(\beta)$ for
the linearisation \eq{schroedinger} of the CNLS example, computed for
$\omega=0.5$ using and interval $x \in [0,10]$. Solid lines
represent asymmetric solutions and dashed lines symmetric ones.
The inserts depict solutions of the linearised equations 
at the first three zeros of $w_1$, both symmetric and antisymmetric, 
which define local bifurcations; and of $w_2$ which define the necessary
conditions for scalar nonlocal bifurcations according to our geometric theory.}
\label{Fig:CNLSbif}
\end{center}
\end{figure}

First let us demonstrate further properties of the structure of
solutions to the linearised problem \eq{schroedinger} by computation of
its even and odd solutions as the parameters vary. Figure \ref{Fig:CNLSbif}
depicts solutions of the constrained linear boundary value problems 
\beq
v_x(0) = 0, \qquad \int_0^X v(x)^2 dx = \mbox{const.}  
\label{autoeven}
\eeq
and 
\beq
v(0)=0, \qquad \int_0^X v(x)^2 dx= \mbox{const.},  
\label{autoodd}
\eeq
for even and odd solutions respectively. Here $X$ is a large
positive constant. At the right-hand boundary point we
can distinguish between solution components that decay with exponential
rate $e^{-\omega x}$ and those which grow with rate
$e^{\omega x}$ by considering the corresponding eigenvectors
in the 
$(v,v')$-plane. 
Hence   
we can define boundary functions
\beq
w_1 = v_x(X) + \omega v(X), \qquad w_2= v_x(X) - \omega v(X), \label{w12def}
\eeq
so that a zero of $w_1$ defines a solution with no exponentially growing
component whereas zeros of $w_2$ define solutions with no component that
decays like $e^{-\omega x}$. Hence according to the above definitions,
$w_1=0$ can be used as a numerical test function for local bifurcations
and $w_2=0$ as a test function for scalar nonlocal bifurcations. 

Specifically Fig.~\ref{Fig:CNLSbif} depicts the results of a numerical
continuation of even and odd solutions to \eq{schroedinger},
satisfying \eq{autoeven} and \eq{autoodd} respectively, for fixed
$\omega$ as $\beta$ is increased from zero. It can be seen that an
alternating sequence of zeros of functions $w_1$ and $w_2$ occurs as
$\beta$ increases. At the values of each of the zeros we plot the mode
shape $v(x)$. Note that each successive pair of zeros corresponds to
the function gaining an extra internal zero. The particular computation
was carried out with $X=10$. For this value it was found that the
$\beta$-values of the depicted zeros of $w_1$ and $w_2$ correspond to
those of the analytic formulae \eq{nonlocal2} and \eq{local2} to
within 5 decimal places. Increasing $X$ resulted in more accuracy, but
an increase in the singularity of the boundary-value problem close to
each zero of $w_1$. 

\begin{figure}[p]
\begin{center}
\parbox{8cm}{\postscript{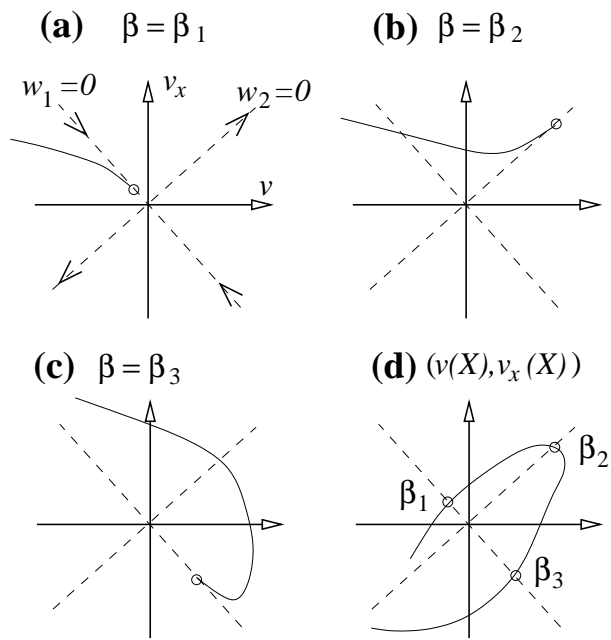}{1.0}}
\caption{Sketch figure illustrating the large-$x$ asymptotics of
even solutions to \eq{schroedinger} as the parameter $\beta$ varies for
fixed $0<\omega<1$. The solution is depicted to the boundary value problem 
\eq{autoeven} up to $x=X$.
Between two $\beta$-values (a),(c) at which condition
$w_1=0$ for local bifurcations occur, there is a $\beta$-value (b) for
which $w_2=0$. Panel (d) sketches the locus of boundary points $(v(X),v_x(X)$
as a function of $\beta$.}  
\label{Fig:clockface}
\end{center}
\end{figure}

For this example we have analytic formulae for the conditions defining
local and scalar nonlocal bifurcations. Hence these computational
results can be interpreted
as developing numerical confidence in our method of detecting them in situations
where analytic formulae do not exist (as in Sec.~\ref{sec:ex2} below).
Also, they provide geometrical insight. Thinking of the phase space $(v,v_x)$,
the conditions for local and scalar nonlocal bifurcations are that
the solution for large $x$ should lie in one of the two eigendirections.
By continuous dependence on initial condition results, then we have
that if there are a succession of local bifurcations with increasing number of
internal zeros upon increasing a parameter,
then the solution at `time' $X$ must rotate in the phase plane. In so doing, 
we can not avoid having a scalar nonlocal bifurcation sandwiched between 
each two successive local bifurcations, see Fig.~\ref{Fig:clockface}.

Next, let us numerically investigate actual solitary wave bifurcations near
the proposed scalar nonlocal bifurcation curve (\ref{nonlocal}). 
First, we consider those curves with $n=1$ and 2. The scalar nonlocal bifurcations
near these curves have been numerically explored in \cite{Ya:97}. 
The results are reproduced in Figs.~\ref{Fig:safig} and \ref{Fig:ssfig}. 
In each figure, solitary waves at three different locations of the
parameter plane are shown: one is close to the local bifurcation curve
(\ref{local}) (dashed line), another one is in the interior, and the third
is close to the theoretical scalar nonlocal bifurcation boundary
(\ref{nonlocal}). Displays of these solitary waves are meant to show the reader
how solitary waves continuously deform from wave and daughter wave structures
as system parameters $\beta$ and $\omega$ vary. 
As we can see, in both cases, scalar nonlocal bifurcations indeed occur
on the theoretical curves (\ref{nonlocal}) (see panel (c) in both figures). 
In addition, the numerical bifurcation boundaries (circles) fall
precisely on the theoretical curves. So for these cases, our 
theory is fully supported by numerics. We have also found similar 
agreement for the case $n=0$ in the nonlocal bifurcation boundary (\ref{nonlocal}).

However, the $n=3$ case is more complicated.  The bifurcation for this
case is shown in Fig.~\ref{Fig:mix}.  The solid line in the parameter
plane (upper left panel) is the theoretical curve (\ref{nonlocal}) for
scalar nonlocal bifurcations.  Circles are numerically detected
nonlocal bifurcation boundaries. Notice that the numerical
boundary falls onto the theoretical curve (\ref{nonlocal}) only in the
lower part. There the nonlocal bifurcation is indeed scalar,
consistent with the geometric argument of Section 2. This can be
confirmed in Fig.~\ref{Fig:mix} (e).  But in the upper and middle
parts, the numerical boundary deviates from the theoretical curve
(\ref{nonlocal}).  The reason turns out to be that, in these parts,
the actual nonlocal bifurcation is {\em not} scalar. Indeed, an inspection
of Fig.~\ref{Fig:mix} (c, d) shows that the bifurcated solitary waves
in these parts are not scalar NLS solitons pieced together. Rather,
they are true vector solitons pieced together.  Thus our 
analysis for scalar nonlocal bifurcations does not apply here.  We
note that near the upper part of the numerical
boundary, the center of the bifurcated wave is a wave and daughter
wave structure with $n=1$ [see Eq. (\ref{local})], and it is flanked
by two single-hump vector solitons on the two sides.  This piecing
together of different vector solitons as a nonlocal consequence of
local bifurcation has been analytically studied in
\cite{Ya:98} before. It was shown there that the boundary for
this type of non-scalar nonlocal bifurcation is precisely the boundary
of local bifurcations (\ref{local}) (here with $n=1$).  This is indeed
the case. When the local-bifurcation boundary (\ref{local}) for $n=1$
is plotted as a dashed curve there, it agrees with the numerical
boundary (circles) very well.  The bifurcation in the middle part of
the parameter region is also non-scalar. It is clear from
Fig.~\ref{Fig:mix} (d) that this bifurcation is somewhere in between
the non-scalar bifurcation of Fig.~\ref{Fig:mix} (c) and the scalar
bifurcation of Fig.~\ref{Fig:mix} (e). In fact, it is appropriate to
consider this middle part of the bifurcation boundary as a transition
between the non-scalar bifurcation in the upper part and the scalar
bifurcation in the lower part.

Numerical searching has revealed that there is no scalar nonlocal
bifurcation observed along the branch corresponding to \eq{nonlocal} with
$n=3$ above the point at which the non-scalar bifurcations start. Thus
the condition \eq{nonlocal} can be at best a necessary condition for
scalar nonlocal bifurcations. 
Why does the bifurcation deviate from scalar here? and where exactly does
this deviation begin? These questions could not be answered by the 
approximate geometric argument in Sec. 2. However, an answer will be revealed 
in a matched asymptotic analysis in the next section. We will show that
the deviation starts where the curve (\ref{nonlocal}) with $n=3$ 
intersects the local bifurcation boundary $\omega=1/s$ of 
$v$-pulses. 

\section{Matched asymptotic theory for
scalar nonlocal bifurcations in the CNLS equations}
\label{sec:NLS_analysis}

To theoretically explain the scalar nonlocal bifurcation results in
the previous section, an analytical theory will now be constructed. 
This theory has three objectives.  The first one is to
prove that the boundaries of scalar nonlocal bifurcations are indeed given by
the condition that the solution of the linear Schr\"odinger equation
(\ref{schroedinger}) has only the purely growing component, i.e.,
Eq. (\ref{nonlocal}).  The second objective is to obtain an analytical
formula for the spacing between the $v$-pulses and the central
$u$-pulse when the parameters are close to the boundary of scalar
nonlocal bifurcations.  The third objective is to determine when 
nonlocal bifurcations can deviate from scalar to non-scalar. 
The technique we will use is similar to the
tail-matching method as developed in \cite{Ya:98} for the construction
of multi-pulse trains, but important modifications need to be made.
Throughout this analysis, we require $\omega<1$ as 
above.

Suppose the ODE system (\ref{u}) and (\ref{v}) allows a solution where
the $u$-component is symmetric and has a dominant pulse in the center
(at $x=0$), while the $v$-component is symmetric or antisymmetric and
has two dominant pulses on the two sides of the $u$-pulse (at $x=\pm
\Delta$). Our main assumption is that the $v$-pulses are
well-separated from the central $u$-pulse, i.e., $\Delta \gg 1$.  Then
we can divide the solution into three regions: (I) the left $v$-pulse
region centered at $x=-\Delta$; (II) the central $u$-pulse region
centered at $x=0$; and (III) the right $v$-pulse region centered at
$x=\Delta$.  Below, we will determine the solutions in each of these
three regions.  Note that midway between region II and I or III, both
the $u$ and $v$ solutions are very small. Thus they are approximately 
governed by the linear parts of Eqs. (\ref{u}) and (\ref{v}), hence these
solutions are linear combinations of purely exponentially growing and
purely exponentially decaying functions to leading order.  
If these tail asymptotics from two adjacent regions can match each other, 
then a solitary wave can be found.  This is the essence of the
tail-matching method.

When the $v$-component is symmetric or antisymmetric, the
tail-matching treatment between regions II and III becomes the same as
that between regions II and I. Thus, we will focus only on matching between
regions I and II.

\begin{figure}[p]
\begin{center}
\parbox{ 12cm}{\postscript{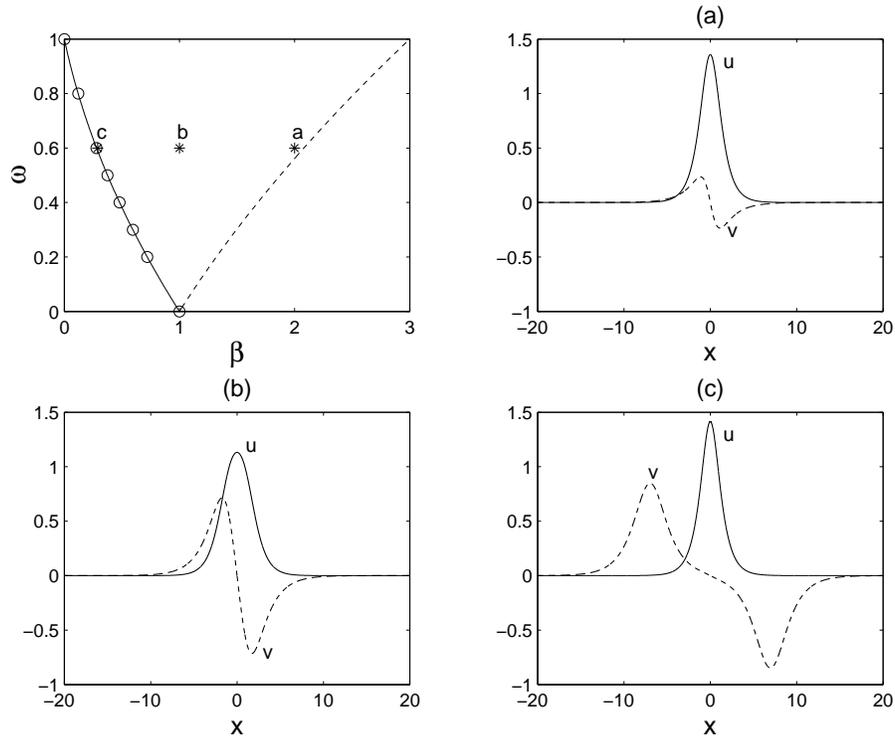}{1.0}}
\caption{Solitary waves in the parameter region (upper left figure)
bounded by the scalar nonlocal 
bifurcation boundary (\ref{nonlocal}) (solid)
and local bifurcation boundary (\ref{local}) (dashed) 
with $n=1$. 
Circles represent the numerical scalar nonlocal bifurcation boundary
for $n=1$ obtained in \protect\cite{Ya:97}. 
Solitary waves at stars marked by letters 'a, b, c' 
in the parameter region are shown with corresponding letters
in the title.  }
\label{Fig:safig}
\end{center}
\end{figure}

\begin{figure}[p]
\begin{center}
\parbox{12cm}{\postscript{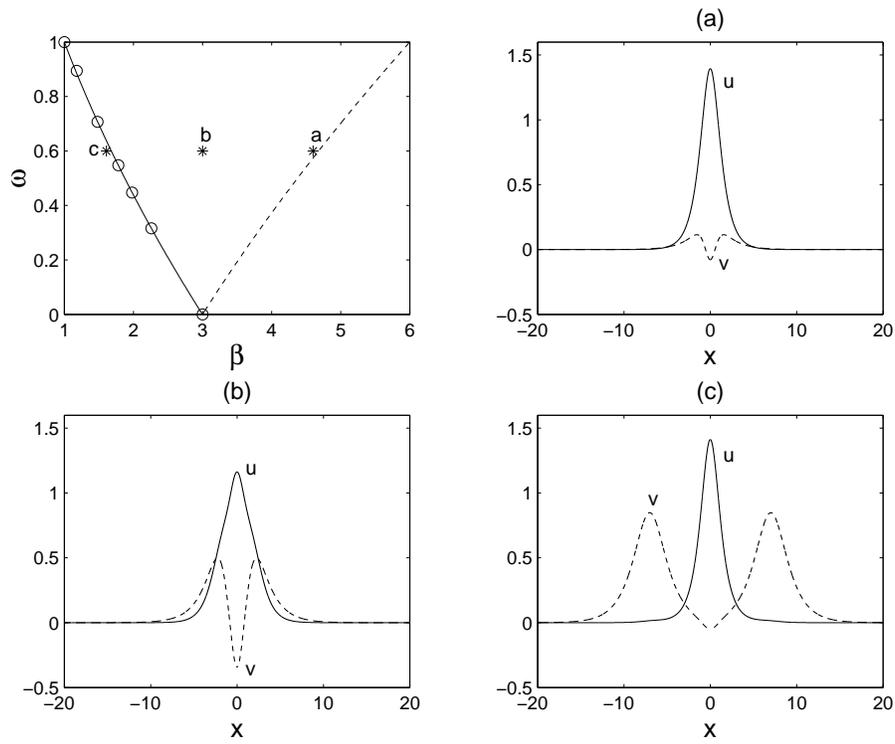}{1.0}}
\caption{Solitary waves in the parameter region bounded by the scalar nonlocal 
bifurcation boundary (\ref{nonlocal}) (solid) and local 
bifurcation boundary (\ref{local}) (dashed)  with $n=2$. 
Circles represent the numerical scalar nonlocal bifurcation boundary
for $n=2$ obtained in \protect\cite{Ya:97}. }
\label{Fig:ssfig}
\end{center}
\end{figure}

\begin{figure}[p]
\begin{center}
\parbox{12cm}{\postscript{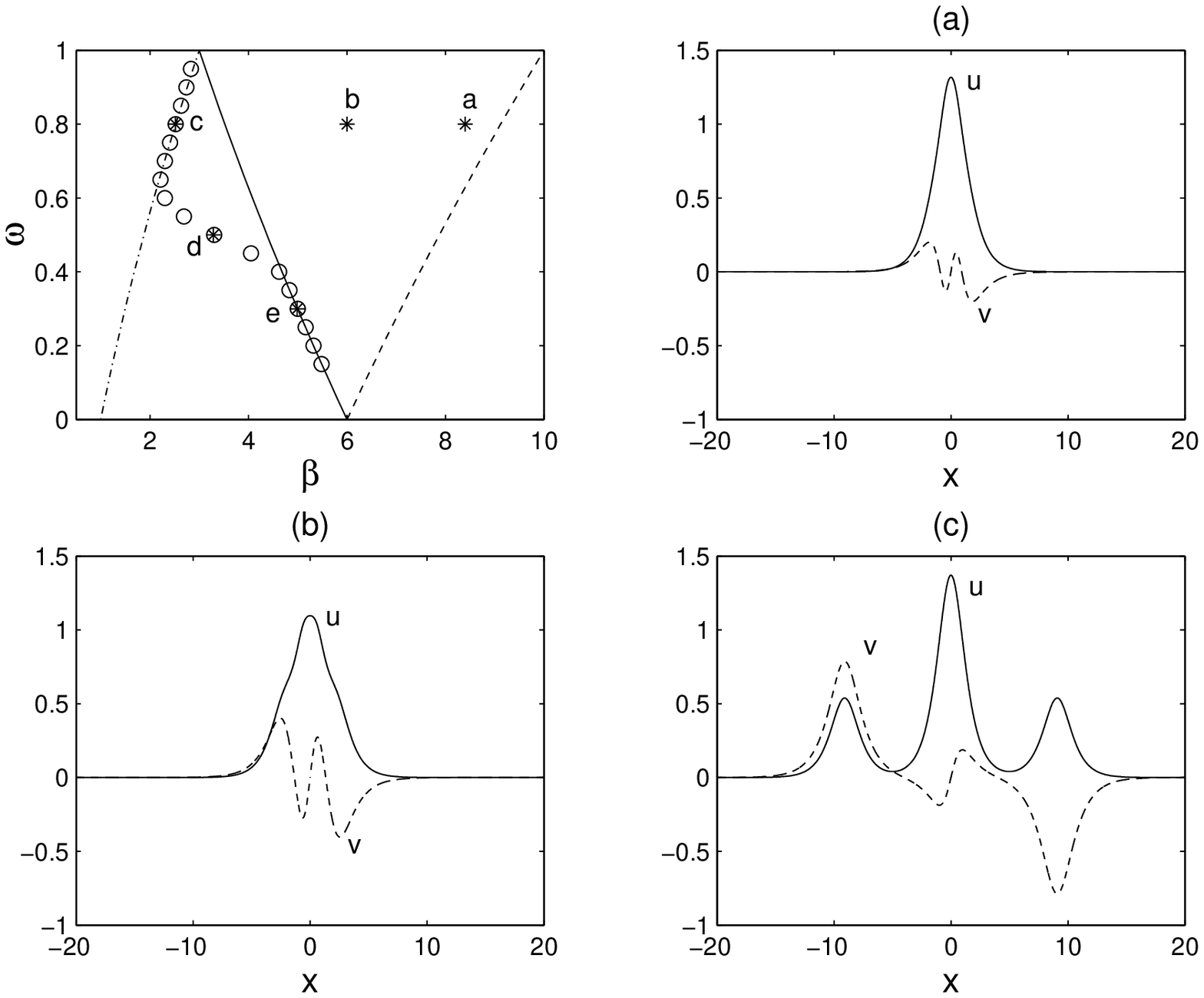}{1.0}}

\parbox{12cm}{\postscript{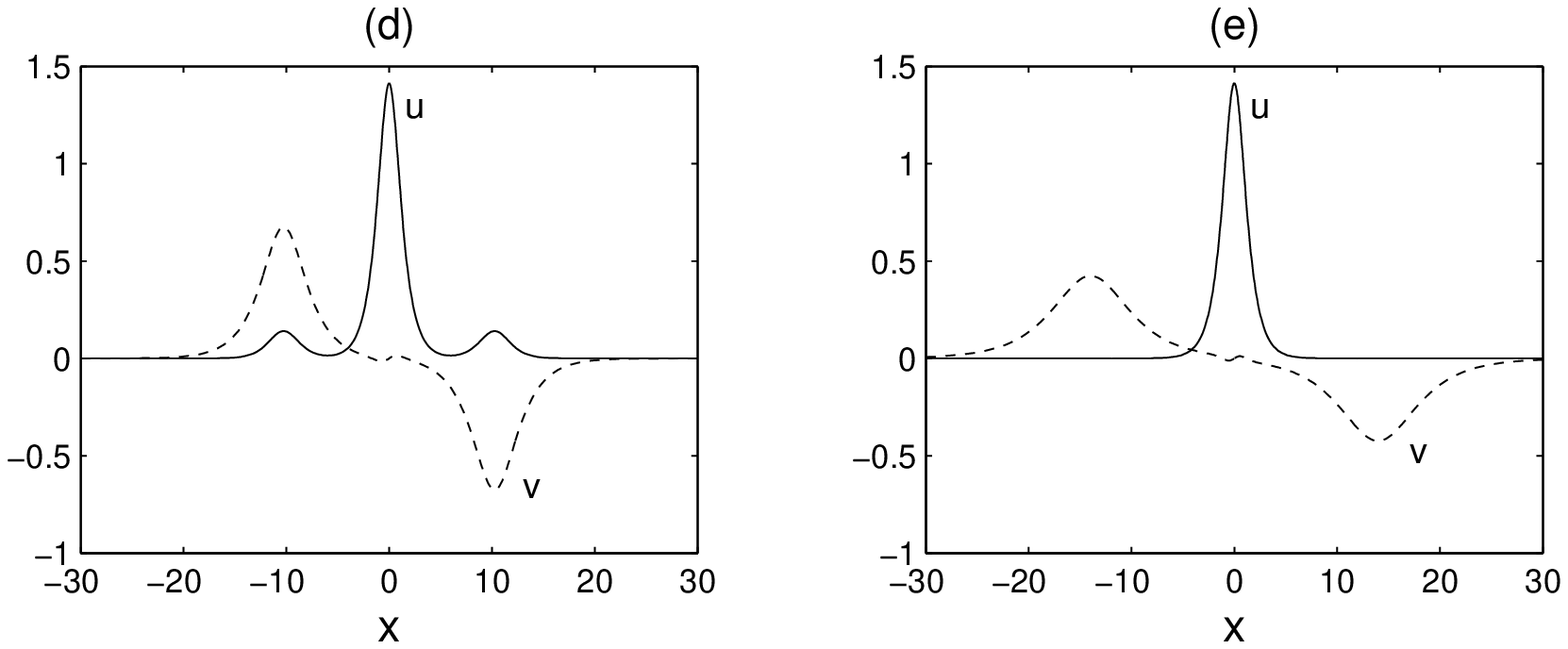}{1.0}}
\caption{Different types of nonlocal bifurcations appearing near the scalar 
nonlocal bifurcation curve (\ref{nonlocal}) with $n=3$ (solid line)
in the coupled NLS system (\ref{u}) and (\ref{v}). 
Circles are numerically obtained nonlocal bifurcation boundaries. 
The lower part of the numerical curve lies on the solid line (\ref{nonlocal}), 
and the bifurcation there is indeed scaler [see (e)]. 
The upper part of the numerical curve lies on the dash-dotted local bifurcation boundary
(\ref{local}) with $n=1$, and the nonlocal bifurcation there is non-scalar
[see (c)]. The nonlocal bifurcation in the middle part of the
numerical curve is somewhere in between scaler and non-scaler [see (d)]. 
The dashed line in the upper left panel is the local bifurcation curve
(\ref{local}) with $n=3$. 
Stars are parameter values where the solitary waves are shown.}
\label{Fig:mix}
\end{center}
\end{figure}

In region I, the solution can be written as

\begin{equation} \label{tildeuvI}
u=\tilde{u}_I, \hspace{0.5cm} v=v_0+\tilde{v}_I, 
\end{equation}
where 
\begin{equation}  \label{tildeuvIb}
v_0=\sqrt{2}\omega \mbox{sech}[\omega (x+\Delta)],  \hspace{0.5cm}
\tilde{u}_I \ll 1, \;\; \tilde{v}_I \ll 1
\end{equation}
(see Figs.~\ref{Fig:safig}(c) and \ref{Fig:ssfig}(c)). 
In new coordinates
\begin{equation} \label{xi}
\xi=x+\Delta, 
\end{equation}
the small $\tilde{u}_I$ component satisfies the linear Schr\"odinger equation
\begin{equation}\label{uIeq}
\tilde{u}_{I\xi\xi}-\tilde{u}_I+\beta v_0^2(\xi)\tilde{u}_I=0
\end{equation}
to leading order. 
To obtain solitary waves, we demand that 
\begin{equation} \label{uIasym1}
\tilde{u}_I(\xi) \longrightarrow 0, \hspace{0.5cm} \xi \rightarrow -\infty. 
\end{equation}
At large positive $\xi$ values, this $\tilde{u}_I$ solution must match 
the tails of the dominant $u$ solution 
\begin{equation}  \label{u0}
u_0(x)=\sqrt{2}\mbox{sech}x, 
\end{equation}
in region II. This matching dictates that the asymptotic behavior
of $\tilde{u}_I$ at large $\xi$ values is
\begin{equation} \label{uIasym2}
\tilde{u}_I(\xi) \longrightarrow 2\sqrt{2}e^{-\Delta}\cdot e^{\xi}, \hspace{0.5cm}
\Delta \gg \xi \gg 1. 
\end{equation}

The linear equation (\ref{uIeq}), together with the boundary
conditions (\ref{uIasym1}) and (\ref{uIasym2}), completely determines
the $\tilde{u}_I$ solution in region I.

Now we determine the small $\tilde{v}_I$ component in region I.  When
Eq. (\ref{tildeuvI}) is substituted into (\ref{v}), and terms of order
$\tilde{v}_I^2$, $\tilde{v}_I^3$ and $\tilde{u}_I^2 \tilde{v}_I$
dropped, we find that, to leading order, $\tilde{v}_I$ satisfies the
following equation
\begin{equation} \label{vIeq}
\tilde{v}_{I\xi\xi}-\omega^2\tilde{v}_{I}+3v_0^2(\xi)\tilde{v}_{I}
=-\beta \tilde{u}_I^2 v_0(\xi). 
\end{equation}
We note that it is important to retain the inhomogeneous term in
Eq. (\ref{vIeq}), as otherwise, that equation with the vanishing
boundary condition at negative infinity would always produce a
localized solution which is impossible to match to the $v$ solution in
region II.  The boundary conditions for solution $\tilde{v}_{I}$ are
\begin{equation} \label{vIasym1}
\tilde{v}_{I} \longrightarrow 0, \hspace{0.5cm} \xi \rightarrow -\infty, 
\end{equation}
and
\begin{equation} \label{vIasym2}
\tilde{v}_{I} \longrightarrow \alpha e^{\omega\xi}+\gamma e^{-\omega \xi}-
\frac{4\sqrt{2}\beta\omega e^{-2\Delta}}{1-\omega} e^{(2-\omega)\xi}, \hspace{0.5cm}
\Delta \gg \xi \gg 1, 
\end{equation}
where $\alpha$ and $\gamma$ are constants. 
The last term in Eq. (\ref{vIasym2}) is contributed from the 
inhomogeneous term of Eq. (\ref{vIeq}). In deriving it, 
the asymptotic behaviors of $\tilde{u}_I$ and $v_0$ solutions
were used [see Eqs. (\ref{tildeuvIb}) and (\ref{uIasym2})]. 

Next, we determine the solutions in region II.  In this region, the
solutions can be written as
\begin{equation} \label{tildeuvII}
u=u_0+\tilde{u}_{II}, \hspace{0.5cm} v=\tilde{v}_{II}, 
\end{equation}
where $u_0(x)$ is given in Eq. (\ref{u0}), and $\tilde{u}_{II},
\tilde{v}_{II} \ll 1$ (see Figs.~\ref{Fig:safig}(c) and
\ref{Fig:ssfig}(c)).  Here we only need to focus on the
$\tilde{v}_{II}$ solution.  This solution satisfies the equation
\begin{equation} \label{vIIeq}
\tilde{v}_{IIxx}-\omega^2 \tilde{v}_{II}+\beta u_0^2(x) \tilde{v}_{II}=0
\end{equation}
to leading order.  The leading asymptotic behavior of this solution at
$x\ll -1$ is
\begin{equation} \label{vIIasym1}
\tilde{v}_{II} \longrightarrow 
\overline{\gamma} \left(e^{-\omega x}+ \delta e^{\omega x}\right), \hspace{0.5cm}
-\Delta \ll x \ll -1, 
\end{equation}
where $\overline{\gamma}$ and $\delta$ are constants.  If we only
consider solitary waves with symmetric or antisymmetric $v$
components, then the above $\tilde{v}_{II}$ solution would have the
same symmetry. This symmetry condition would uniquely determine the
coefficient $\delta$.  The constant $\overline{\gamma}$ is selected by
the condition that the tail asymptotics of $\tilde{v}_{II}$ solution
at $x \ll -1$ in region II must match the $v$ solution
(\ref{tildeuvI}) at $\xi \gg 1$ in region I.  This matching gives 
$\overline{\gamma}$ as $=(2\sqrt{2}\omega+\gamma) e^{-\omega \Delta}$. 
Recall that $\tilde{v}_I \ll 1$, and thus $\gamma \ll 1$. 
As a result, to the leading order, 
\begin{equation}
\overline{\gamma}=2\sqrt{2}\omega e^{-\omega \Delta}. 
\end{equation}
This matching also gives the relation
\begin{equation} \label{alpha}
\alpha=\overline{\gamma}\delta \hspace{0.03cm} e^{-\omega \Delta}
=2\sqrt{2}\omega \delta \hspace{0.03cm} e^{-2\omega \Delta}.
\end{equation}
One may wonder why the third term in the 
$\tilde{v}_I$ asymptotics (\ref{vIasym2})
is not matched by $\tilde{v}_{II}$ asymptotics 
(\ref{vIIasym1}). In fact, there is a smaller term
in the $\tilde{v}_{II}$ solution which is proportional to
$e^{(2-\omega)x}$.  This term arises due to the product of $u_0^2$ and
the $e^{-\omega x}$ component in the leading $\tilde{v}_{II}$ solution
(\ref{vIIasym1}) [see Eq. (\ref{vIIeq})].  
One can check that this term will exactly match the
third term in the $\tilde{v}_I$ 
asymptotics (\ref{vIasym2}). So there is no
contradiction here.  But this is a minor issue which is not critical
to our analysis.

Now we are in a position to derive a formula for the spacing $\Delta$
between the $v$-pulses and the middle $u$-pulse.  This formula comes
from the solvability condition for the $\tilde{v}_{I}$ equation
(\ref{vIeq}) together with the boundary conditions (\ref{vIasym1}),
(\ref{vIasym2}) and (\ref{alpha}).  It is noted that Eq. (\ref{vIeq})
is self-adjoint, and it has a localized homogeneous solution
$v_{0\xi}$ due to the spatial translation invariance of the ODE
equation (\ref{v}).  Calculating the integrals of products between
$v_{0\xi}$ and the two sides of Eq. (\ref{vIeq}) from $-\infty$ to
$y$, and integrating by parts, we get
\begin{equation}
\int_{-\infty}^{y}-\beta \tilde{u}_I^2 v_0 v_{0\xi}d\xi
=\left. \left(\tilde{v}_{I\xi}v_{0\xi}-\tilde{v}_{I}v_{0\xi\xi}\right)\right|_{-\infty}
  ^{y}. 
\end{equation}
When $\Delta \gg y\gg 1$, substituting the boundary conditions (\ref{vIasym1}),
(\ref{vIasym2}) and (\ref{alpha}) into the above equation, we find
that
\begin{equation} \label{integral}
\int_{-\infty}^{y}-\beta \tilde{u}_I^2 v_0 v_{0\xi}d\xi
\longrightarrow 
-16\delta \omega^4 e^{-2\omega\Delta}+\frac{32\beta \omega^3 e^{-2\Delta}}{1-\omega}
e^{2(1-\omega)y}, \hspace{0.5cm} y \rightarrow \infty.
\end{equation}
The above equation is the leading two-term expansion for the
integral on its left-hand side.  When $y$ approaches infinity, this
integral diverges. But we can separate this divergent part from the
rest of the integral.  Notice that
\begin{equation}
\frac{32\beta \omega^3 e^{-2\Delta}}{1-\omega}e^{2(1-\omega)y}
=\int_{-\infty}^{y} 64\beta\omega^3e^{-2\Delta} e^{2(1-\omega)\xi} d\xi. 
\end{equation}
Thus Eq. (\ref{integral}) can be rewritten as
\begin{equation} \label{formula0}
\beta \int_{-\infty}^{\infty} \left[\tilde{u}_I^2(v_0^2)_{\xi}+128\omega^3
e^{-2\Delta} e^{2(1-\omega)\xi}\right] d\xi = 32\delta \omega^4e^{-2\omega\Delta}. 
\end{equation}
The integral above is no longer divergent. 
In fact, one can use the asymptotic relations
(\ref{uIasym1}) and (\ref{uIasym2}) to check that the integrand in
that integral approaches zero exponentially as $|x|$ goes to
infinity.

Eq. (\ref{formula0}) gives a formula for spacing $\Delta$ when the
system parameters $\beta$ and $\omega$ are specified. This formula can
actually be made more explicit as follows. Recall that function
$\tilde{u}_I$ is determined by Eq. (\ref{uIeq}) and boundary
conditions (\ref{uIasym1}) and (\ref{uIasym2}). Under the notation
\begin{equation} \label{notation}
\tilde{u}_I(\xi)=2\sqrt{2}\: e^{-\Delta}\phi(\xi),
\end{equation}
function $\phi(\xi)$ is then uniquely specified by the following equation
and boundary conditions
\begin{equation} \label{phi}
\phi_{\xi\xi}-\phi+\beta v_0^2(\xi)\phi=0, 
\end{equation}
\begin{equation} \label{phibc}
\phi(\xi) \longrightarrow 
\left\{ \begin{array}{ll} 0, & \xi \rightarrow -\infty, \\
e^{\xi}, & \xi \rightarrow \infty,
\end{array} \right. 
\end{equation}
where $v_0(\xi)$ is given by Eqs. (\ref{tildeuvIb}) and (\ref{xi}).
Under these notations, formula (\ref{formula0}) simplifies as
\begin{equation} \label{formula}
e^{-2(1-\omega)\Delta}=\frac{4\omega^4 \delta}{\beta I},
\end{equation}
where $I$ is the integral 
\begin{equation}
I=\int_{-\infty}^\infty \left[ \phi^2(v_0^2)_{\xi}+16\omega^3e^{2(1-\omega)\xi}\right]d\xi.
\end{equation}
Recall that the constant $\delta$ is defined by
Eqs. (\ref{vIIeq}) and (\ref{vIIasym1}). To be more explicit, $\delta$
is defined by 
\begin{equation} \label{psi}
\psi_{xx}-\omega^2\psi+\beta u_0^2(x)\psi=0, 
\end{equation}
and 
\begin{equation} \label{psibc}
\psi(x) \longrightarrow e^{-\omega x}+\delta e^{\omega x}, \hspace{0.5cm} x \rightarrow -\infty.
\end{equation}
In other words, $\delta$ is the coefficient of the purely decaying
component of the Schr\"odinger equation (\ref{psi}) at $x=-\infty$.
Function $\psi$'s boundary condition at $x=\infty$ is provided by the
symmetry of the $v$ component in the solitary wave we are
seeking. Since we are focusing on symmetric and antisymmetric $v$
components, function $\psi$ would have the same symmetry.  This
symmetry helps to uniquely determine the $\delta$ coefficient in the
above linear problem.

\begin{figure}[p]
\begin{center}
\parbox{6cm}{\postscript{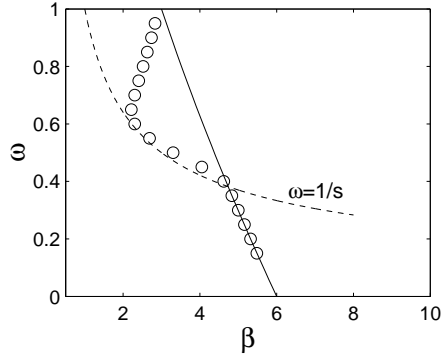}{1.0}}

\caption{Intersection between the dashed curve $\omega=1/s$ [see Eq. (\ref{sbeta})]
defining local bifurcations of daughter-$u$ solutions
from $v$-pulses (with $n=0$) 
 and the solid curve defining the necessary condition (\ref{nonlocal}) 
for scalar nonlocal bifurcations
(with $n=3$). Circles show the numerical results for nonlocal bifurcations
reproduced from Fig.~\ref{Fig:mix}.}  
\label{mix:copy}
\end{center}
\end{figure}

Formula (\ref{formula}) is the key result of this section.  It
explicitly gives the expression for the spacing $\Delta$ in solitary
waves bifurcating 
from scalar nonlocal bifurcations. Several observations quickly follow
from this formula.  First, solitary waves from scalar nonlocal bifurcations
exist only when parameters $\delta$ and $I$ have the same sign.
Second, when $\delta=0$, $\Delta$ goes to infinity. Thus this is a boundary
of scalar nonlocal bifurcations. This condition is precisely the one proposed
in Sec. 2. For the coupled NLS system, $\delta=0$ 
on the curves (\ref{nonlocal}). One may notice from formula
(\ref{formula}) that $\Delta$ also goes to infinity when $I=\infty$. 
However, $I=\infty$ does not correspond to a boundary of scalar
nonlocal bifurcations. The reason is as follows. 
For the coupled NLS system, $I=\infty$ on the local bifurcation boundaries of
pure $v$-pulses: 
\begin{equation} \label{Iinfty}
\omega=\frac{1}{s-n},
\end{equation}
where $n$ is an integer and $0\le n< s$. This is because on these boundaries, 
the solution $\phi$ of Eq. (\ref{phi}) which satisfies the zero boundary 
condition at $\xi=-\infty$ [see Eq. (\ref{phibc})] is always
localized. Thus in order for it to satisfy the boundary condition
(\ref{phibc}) at $\xi=\infty$, $\phi$ must be infinitely large. 
Hence $I=\infty$. The above fact applies to the solution 
$\tilde{u}_I$ of Eq. (\ref{uIeq}) as well: on the local bifurcation boundary
(\ref{Iinfty}), solution $\tilde{u}_I$ satisfying boundary conditions
(\ref{uIasym1}) and (\ref{uIasym2}) is infinitely large. When this happens, 
our original assumption $\tilde{u}_I \ll 1$ for scalar nonlocal bifurcations
breaks down. Hence if there is a nonlocal bifurcation here at all, it would 
not be scalar: the pulses on the two wings would be true vector solitons. 
Thus $I=\infty$ does not give a boundary of scalar nonlocal bifurcations. 
Consequently, a scalar nonlocal bifurcation boundary is given entirely
by the condition $\delta=0$, which is our previous condition. 
An interesting and subtle issue is: what if curves $\delta=0$ and $I=\infty$
intersect? As we have discussed above, when $I=\infty$, the nonlocal 
bifurcation (if there is one) becomes non-scalar. Thus at the locus of
$I=\infty$ and $\delta=0$, the nonlocal bifurcation could turn from scalar 
to non-scalar. Then the bifurcation boundary would deviate
from the scalar bifurcation curve $\delta=0$ at the intersection. 
This phenomenon could, and does, happen. In fact, Fig. \ref{Fig:mix}
gives a good example. Let us reproduce Fig. \ref{Fig:mix}'s scalar
bifurcation curve [i.e., (\ref{nonlocal}) with $n=3$] and the true 
bifurcation boundary in Fig. \ref{mix:copy} below (solid line and circles). 
On top of it, we plot the local bifurcation curve (\ref{Iinfty}) of $v$-pulses
with $n=0$ (dashed line). We see that the intersection between these two curves
is precisely where bifurcation turns from scalar to non-scalar, thus
deviation between the numerical bifurcation boundary and the scalar 
bifurcation boundary starts there. 
This example and the matched asymptotic analysis above tell us that the
previous condition for scalar nonlocal bifurcations is 
a necessary but not sufficient condition. When local bifurcation curves
of $v$-pulses intersect with these necessary-condition curves, non-scalar 
bifurcations could start, thus scalar bifurcations on part of the
necessary-condition curves will not materialize. 

\begin{figure}[p]
\begin{center}
\parbox{5cm}{\postscript{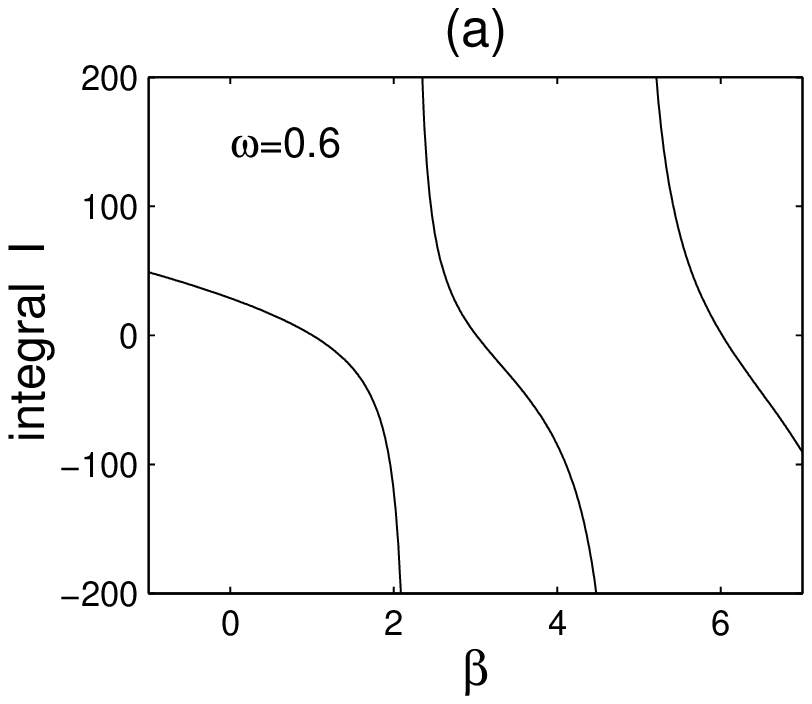}{1.0}}
\hspace{0.4cm}
\parbox{5cm}{\postscript{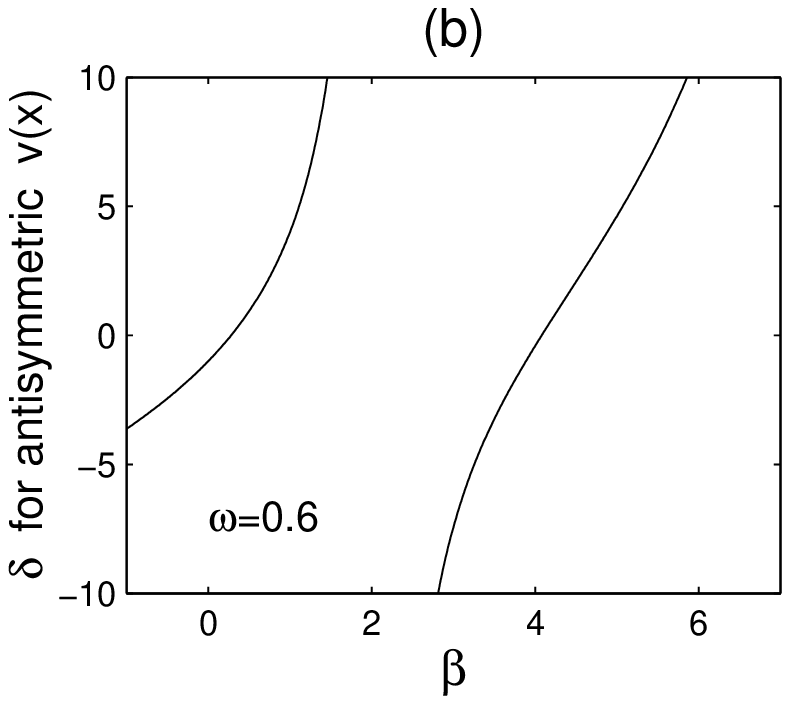}{1.0}}
\parbox{5cm}{\postscript{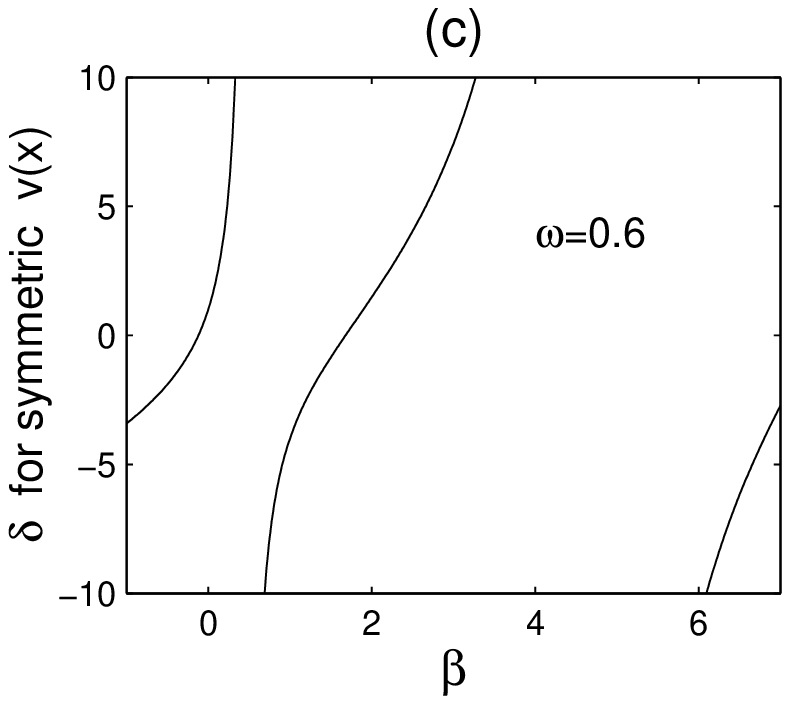}{1.0}}
\caption{
Parameter values $\delta$ and $I$ at $\omega=0.6$ and
various $\beta$ values.
(a) the integral $I$ values; (b) coefficient $\delta$ 
for anti-symmetric $v$ components; 
(c) $\delta$ for symmetric $v$ components. 
}
\label{deltaI}
\end{center}
\end{figure}

The analytical expressions for $\delta$ and $I$ may be possible to
obtain, as the linear Schr\"odinger equations (\ref{phi}) and
(\ref{psi}) can be solved using hyper-geometric functions (see
\cite{LaLi:77} and Sec. 3).  But such expressions would be very
complex. For practical purposes, it is preferable to determine them
numerically. For illustration purposes, we select $\omega=0.6$, and
show the numerical values of these quantities in Fig.~\ref{deltaI} at
various $\beta$ values ranging from $-1$ to 7.  In Fig.~\ref{deltaI}a,
the integral $I$ is shown. This integral is independent of the
symmetry of the $v$ solution.  In Fig.~\ref{deltaI}(b, c), the
coefficients $\delta$ for anti-symmetric and symmetric $v$ solutions
are shown respectively. This figure is helpful in revealing on which
side of the scalar nonlocal bifurcation boundary solitary waves can be
expected. For instance, at $\omega=0.6$, $\delta=0$ when $\beta=0.28$
and the $v(x)$ component is anti-symmetric.  On the left-hand side of
this $\beta$ value, $\delta>0$ and $I>0$, thus solitary waves with
symmetric $u$ component and anti-symmetric $v$ component can be
expected. But on the right-hand side of this $\beta$ value, $\delta<0$
and $I>0$, thus no such waves can be found.  These predictions
completely agree with the numerics shown in the previous section.
Similar agreement is found near other scalar nonlocal bifurcation curves 
as well. 

As $\omega \rightarrow \omega_n^{NLB}(\beta)$, 
$\delta \rightarrow 0$, thus $\Delta \rightarrow \infty$ according to 
formula (\ref{formula}). Below, we derive the asymptotic formula for 
$\Delta$ when $\omega \rightarrow \omega_n^{NLB}(\beta)$, or
equivalently, $\beta \rightarrow \beta_n^{NLB}(\omega)$ [see Eq. (\ref{nonlocal2})]. 
The latter limit will be adopted in the following derivation as it is 
a little more convenient. 

When $\beta \rightarrow \beta_n^{NLB}(\omega)$, the integral $I$
approaches a finite constant value $I(\omega, \beta_n^{NLB})$, while
$\delta$ goes to zero. Obviously, the asymptotic formula 
for $\Delta$ crucially depends on the asymptotic formula of $\delta$. 
We determine $\delta$'s asymptotic formula by regular perturbation methods below. 
Equation (\ref{psi}) can be rewritten as
\begin{equation} \label{psipert}
\psi_{xx}-\omega^2\psi+\beta_n^{NLB} u_0^2\psi=-\epsilon u_0^2 \psi, 
\end{equation}
where $\epsilon=\beta-\beta_n^{NLB}$. 
When $\epsilon=0$, this equation has an unbounded solution 
\begin{equation}
\psi_{n0}(x)=\left\{\begin{array}{ll}\frac{1}{g_n}\hat{v}_1(x), & n=2n_1, \\
-\frac{1}{g_n}\hat{v}_2(x), & n=2n_2+1, \end{array}\right.
\end{equation}
where functions $\hat{v}_{1,2}(x)$ and constant $g_n$ are given 
by Eqs. (\ref{v1n}), (\ref{v2n}) and (\ref{gn}). 
The asymptotic behavior of this solution is that
\begin{equation} \label{psin0bc}
\psi_{n0}(x) \longrightarrow e^{-\omega x} + o\left(e^{\omega x}\right), 
\hspace{0.6cm} x \rightarrow -\infty
\end{equation}
[see Eqs. (\ref{vasym})]. 
When $\epsilon$ is small, 
solution $\psi$ can be expanded into a regular perturbation series
\begin{equation}
\psi(x,\beta,\omega)=\psi_{n0}(x,\beta_n^{NLB},\omega)+\epsilon \psi_{n1}(x,\beta_n^{NLB},\omega)
+O(\epsilon^2). 
\end{equation}
When this expansion is substituted into Eq. (\ref{psi}), 
at order $\epsilon$, we find that function $\psi_{n1}$ satisfies the equation
\begin{equation} \label{psin1}
\psi_{n1xx}-\omega^2 \psi_{n1}+\beta_n^{NLB} u_0^2 \psi_{n1}=-u_0^2\psi_{n0}.
\end{equation}
Recalling that the boundary conditions for $\psi(x)$ and $\psi_{n0}(x)$
are given by Eqs. (\ref{psibc}) and (\ref{psin0bc}), 
thus the boundary condition for function $\psi_{n1}$ is
\begin{equation} \label{psin1bc}
\psi_{n1}(x) \longrightarrow \frac{\delta}{\epsilon} e^{\omega x}, \hspace{1cm}
x \rightarrow -\infty.
\end{equation}  
The linear operator on the left hand side of Eq. (\ref{psin1}) is 
self-adjoint. In addition, $\psi_{n0}(x)$ is a homogeneous solution. 
Calculating the inner product between $\psi_{n0}$ and the inhomogeneous
term of Eq. (\ref{psin1}), we readily find that
\begin{equation}
\int_{-\infty}^{\infty}u_0^2\psi_{n0}^2 dx
=\left. \left(\psi_{n0x}\psi_{n1}-\psi_{n0}\psi_{n1x}\right) \right|_{-\infty}^{\infty}.
\end{equation}
Substituting the boundary conditions 
(\ref{psin0bc}) and (\ref{psin1bc}) into the above relation, 
we find that for both $\psi(x)$ symmetric and anti-symmetric cases, 
the constant $\delta$ is given by the asymptotic formula
\begin{equation}
\delta=\frac{\beta-\beta_n^{NLB}}{4\omega}\int_{-\infty}^{\infty}
u_0^2\psi_{n0}^2 dx + O\left((\beta-\beta_n^{NLB})^2\right).
\end{equation}
We have compared this formula with the numerical values of $\delta$
as displayed in Fig.~\ref{deltaI}(b,c). The slope 
$\int_{-\infty}^{\infty}u_0^2\psi_{n0}^2 dx/4\omega$
predicted by this
formula is in excellent agreement with the numerical slope at 
$\beta=\beta_n^{NLB}$. 
When this formula is substituted into Eq. (\ref{formula}), we finally obtain
the leading two-term 
asymptotic expansion for spacing function $\Delta$ as
\begin{equation} \label{Deltaexpansion}
\Delta=\frac{1}{2(\omega-1)}\left\{\ln(\beta-\beta_n^{NLB})+\ln K+
O(\beta-\beta_n^{NLB})\right\},
\end{equation}
where the constant $K$ is
\begin{equation}
K=K_n(\omega)=\frac{\omega^3}{\beta_n^{NLB} I(\omega, \beta_n^{NLB})}
\int_{-\infty}^{\infty}u_0^2\psi_{n0}^2 dx.
\end{equation}

\begin{figure}[p]
\begin{center}
\parbox{14cm}{\postscript{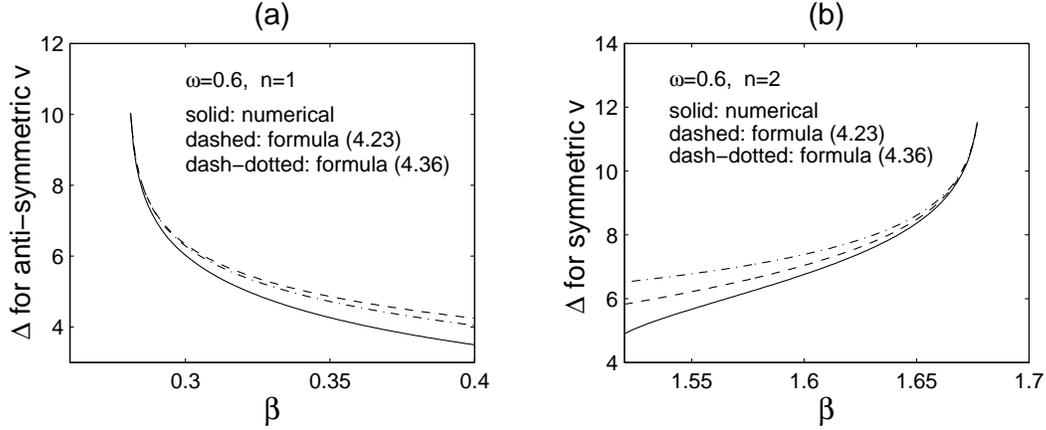}{1.0}}
\caption{Comparison between the analytical formula (\ref{formula}), its
leading two-term 
asymptotic expansion (\ref{Deltaexpansion}) and numerical values 
for spacing $\Delta$ at $\omega=0.6$ and various $\beta$ values. 
(a) comparison near the scalar nonlocal bifurcation boundary (\ref{nonlocal}) with $n=1$; 
(b) comparison near the scalar nonlocal bifurcation boundary (\ref{nonlocal}) with $n=2$. 
\label{spacing} }
\end{center}
\end{figure}

Next, we make quantitative comparisons between the spacing formulas
(\ref{formula}), its leading two-term expansion (\ref{Deltaexpansion}) 
and numerics near the scalar nonlocal bifurcation boundaries
(\ref{nonlocal}) with $n=1$ and 2 at $\omega=0.6$ and various $\beta$
values (see Fig.~\ref{Fig:fig1}).  We remind the reader that
$\omega=1-s$ is a boundary for anti-symmetric $v$ components, and
$\omega=2-s$ is a boundary for symmetric $v$ components.  At
$\omega=0.6$, the boundary point is at $\beta_1^{NLB}=0.28$ for the former
case, and is at $\beta_2^{NLB}=1.68$ for the latter case.  The analytical
spacings from formula (\ref{formula}) and its two-term
asymptotic expansion (\ref{Deltaexpansion})
are shown as dashed and dash-dotted lines in Fig.~\ref{spacing}(a, b)
for these two cases respectively
We have also numerically determined
the spacings between the $v$-pulses and the central $u$-pulse in the
exact solitary waves.  These numerical values are shown as solid lines
in Fig.~\ref{spacing}(a, b).  We see that, when the separation
$\Delta$ is large, the formula (\ref{formula}) and its asymptotic form 
(\ref{Deltaexpansion}) agree with the numerical values perfectly.


\section{Example 2: Saturable nonlinearity}
\label{sec:ex2}

As a second example we take the coupled nonlinear Schr\"{o}dinger system
studied by Ostrovskaya and Kivshar \cite{OsKi:99} 
\begin{eqnarray}
i U_t + U_{xx} 	     + U \frac{|U|^2+|V|^2}{1+s(|U|^2+|V|^2)} & = & 0
\label{elenaPDE1}\\
iV_t +  V_{xx} + V \frac{|U|^2+|V|^2}{1+s(|U|^2+|V|^2)} & = & 0.
\label{elenaPDE2}
\end{eqnarray}
This dimensionless model arises after scaling of a model for the incoherent interaction
between two linearly polarized optical beams in a biased photorefractive medium.  
Here $s$ ($0<s<1$) is an effective saturation parameter,
representing the photorefractive effects. 
This system is significant because it was shown in \cite{Osetal:99}
that multi-humped stationary pulses may be stable solutions, a result
that explains the experimental observations of
\cite{MiSeCh:98}. In the limit $s \to 0$, this system reduces to 
the Manakov equations. 

\begin{figure}[p]
\begin{center}
\parbox{8cm}{\postscript{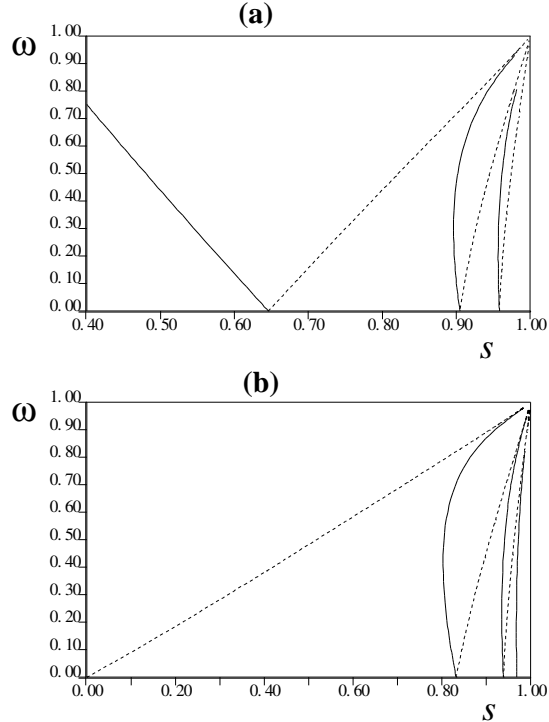}{1.0}}
\caption{Curves of local (dashed) and scalar nonlocal (solid) 
bifurcations from the pure $u$-pulse $u_h$ of 
\eq{elena1}, \eq{elena2} obtained by numerically solving
the linearised equation \eq{elena_lin} and 
requiring the solution to have purely decaying or growing asymptotics at infinity.
(a) the $v$-component is even; (b) the $v$-component is odd. 
Only the first three
curves of each type is presented.}
\label{Fig:Elena1}
\end{center}
\end{figure}

The solitons in this system are of the form
\begin{equation}
U(x,t)=e^{it}u(x), \hspace{0.5cm} V(x,t)=e^{i\omega^2 t}v(x), 
\end{equation}
where $u$ and $v$ are real functions satisfying the following ODEs: 
\begin{eqnarray}
u_{xx}  -u 	     + u \frac{u^2+v^2}{1+s(u^2+v^2)} & = & 0,
\label{elena1}\\
v_{xx}  -\omega^2 v + v \frac{u^2+v^2}{1+s(u^2+v^2)} & = & 0.
\label{elena2}
\end{eqnarray}
Looking for single-component pulses ($u=0$, or $v=0$) we obtain simple planar 
equations, which can be shown by phase plane techniques to possess
symmetric homoclinic orbits $\pm u_h$, $\pm v_h$. Unlike the previous
example, we know of no closed form expressions for these solutions
other than at $s=0$. So we turn straight away to numerical
methods. Once again we restrict to $0<\omega<1$ and look for
local and scalar nonlocal bifurcations from the pulse $u_h$.
Figure \ref{Fig:Elena1} shows the analogue of Fig.~\ref{Fig:fig1} for this 
example where the bifurcation boundaries
were obtained by numerically imposing the boundary conditions 
for local bifurcation and for scalar nonlocal bifurcation on
the linearised equation 
\begin{equation}
v_{xx}-\omega^2 v + \frac{ u_h^2(x) v}{1+ s u_h^2(x)}=0.
\label{elena_lin}
\end{equation}

\begin{figure}[p]
\begin{center}
\parbox{8cm}{\postscript{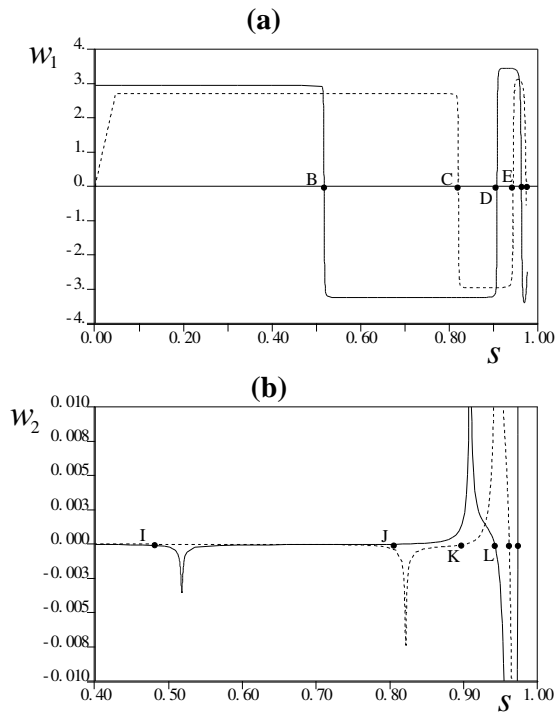}{1.0}}
\caption{Curves of (a) $w_1(s)$ and (b) $w_2(s)$ 
defining respectively local and scalar nonlocal bifurcations
according to the theory, for the model
\eq{elena1}, \eq{elena2} at $\omega=0.5$. Dashed
lines correspond to modes with even symmetry and solid
lines to odd-symmetric modes. 
The labeled zeros of each function correspond to mode
shapes for $v(x)$ that are {\em qualitatively} similar
(i.e. having the same structure of zeros) as the 
corresponding panels of Fig.~\ref{Fig:CNLSbif}. }
\label{Fig:Elenabif1}
\end{center}
\end{figure}

Figure \ref{Fig:Elenabif1} depicts the corresponding 
graphs of $w_1(s)$ and $w_2(s)$ for fixed $\omega=0.5$, where
$w_1$ and $w_2$ were defined in \eq{w12def}.
Note that we have qualitatively the same structure as
in Fig.~\ref{Fig:CNLSbif} in that the mode shapes
$v(x)$ gain increasingly many internal zeros as the parameter
increases to some limit. Moreover, between each pair of
local bifurcations of a given symmetry type there is a scalar nonlocal
bifurcation, and vice-versa.
Here it would seem that the limit 
$s \to 1$ plays the same qualitative role as $\beta \to \infty$
in the previous model, and we conjecture that there are infinitely
many local and scalar nonlocal bifurcations as we approach this limit.
This is bourn out by the results in Fig.~\ref{Fig:Elena1} which show
that the curves in the $(s,\omega)$-plane defined by zeros
of $w_1=0$ and $w_2=0$ become increasingly steep as $s \to 1$.
It would be fair to conjecture from the numerics that all curves (apart
from the first symmetric scalar nonlocal bifurcation) approach the
point $(\omega,s)=(1,1)$. There are some numerical difficulties
in computing the nonlocal bifurcation curves up to this
point, because the pulse $u_h$ becomes infinitely broad as $s \to 1$, and
hence the solution $v(x)$ of the linearised problem which grows 
exponentially at infinity must be continually 
rescaled to avoid the solution becoming arbitrarily large.

Now the presence of local bifurcations has been found before
in this system; see Fig. 1 in \cite{OsKi:99} where local bifurcations
are found via zeros of a certain integral as a function of $\omega$ for
several different values of $s$. Here we have presented a simple
procedure for automatically detecting these local bifurcations
as a function of all system parameters. As far as we know these
curves of local bifurcations are not given by closed form analytic
expressions as they were in the previous example.

\begin{figure}[p]
\begin{center}
\parbox{10cm}{\postscript{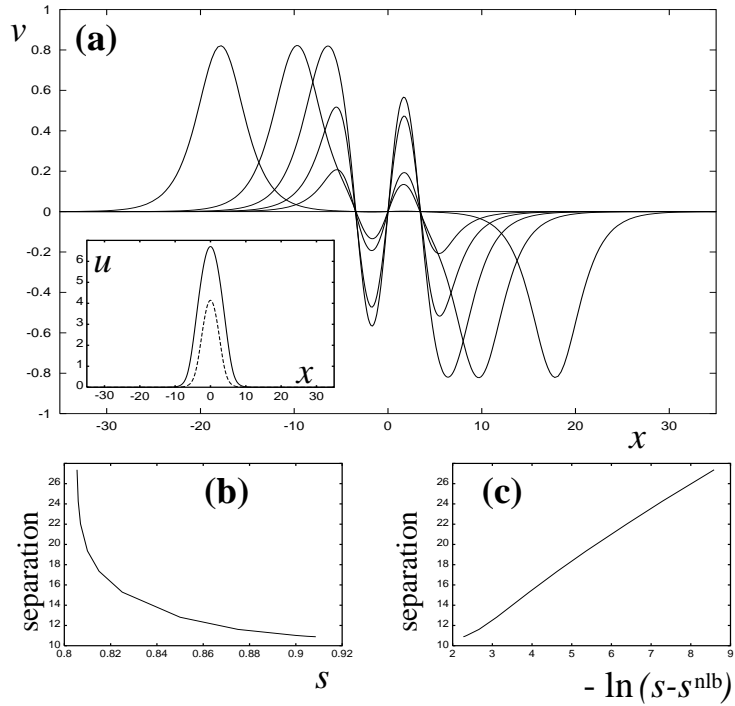}{1.0}}
\caption{(a) The $v$-component of vector solitons
in \eq{elena1}, \eq{elena2} for $\omega=0.5$
and $s=0.9075 (\approx s^{LB}), 0.9, 0.85, 0.81,
0.8055 (\approx s^{NLB})$. The inset depicts
the $u$-component of the vector soliton at $s^{LB}$
(larger peak) and $s^{NLB}$ (smaller peak). In (b) the
separation between the two maxima of $|v|$ 
is plotted as a function of $s$, and
(c) plots the same data on a logarithmic scale
where the more precise value $s^{NLB}=0.8053134$
is used.}
\label{Fig:Elenasep}
\end{center}
\end{figure}

The question remains
whether the curves which have been putatively called scalar nonlocal bifurcations
really represent that for the full equations. 
Figure \ref{Fig:Elenasep} depicts numerical solutions of the fully
coupled nonlinear equations for fixed $\omega=0.5$ in a parameter
region between the second local and first scalar nonlocal symmetric
bifurcations. A one-parameter family of vector solitons is computed
which are born at the local bifurcation. It is found that the solution
branch terminates at the value $s\approx 0.8055$ which agrees to
four decimal places with the scalar nonlocal bifurcation boundary
determined from our geometric argument of Sec. 2. 
Inspection of solitary waves close to this boundary indicates that
a scalar nonlocal bifurcation indeed occurs here. Note from Fig. 
\ref{Fig:Elenasep}(c) panel that the separation varies logarithmicly 
with the variation of the parameter away from the nonlocal bifurcation
boundary, as in the first example [see Eq. (\ref{Deltaexpansion})]. 

\begin{figure}[p]
\begin{center}
\parbox{8cm}{\postscript{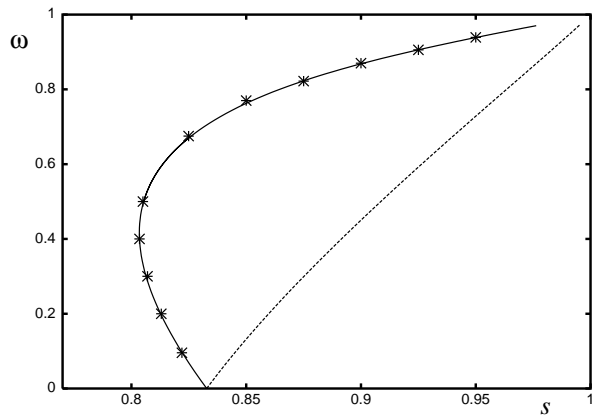}{1.0}}
\caption{A scalar nonlocal bifurcation boundary for
the system \eq{elena1}, \eq{elena2}. The stars
represent scalar nonlocal bifurcation boundaries that
were obtained by solving the full nonlinear equations.
The solid line was obtained from the condition $w_2=0$ for
the linearised 
equation \eq{elena_lin}. The dashed line is a similarly obtained
curve of local bifurcations from which these particular vector
solitons are born.} 
\label{Fig:asymg}
\end{center}
\end{figure}

Finally, Fig.~\ref{Fig:asymg} indicates the results
of a series of similar one-parameter continuations (for either fixed
$s$ or fixed $\omega$) starting from the same curve of local
bifurcations. Observe the agreement between the termination of these
branches and the theoretical 
scalar nonlocal bifurcation boundary. Indeed, in each run
the final orbit was found to be a pure $u$-pulse flanked symmetrically
by two $v$-pulses, qualitatively the same as in
Fig.~\ref{Fig:Elenasep}. Similar results starting from each of the 
local bifurcation boundaries computed in Fig.~\ref{Fig:Elena1} have found
that the bifurcated branches all continue smoothly up to the 
theoretical scalar nonlocal bifurcation
curves. Thus, in this second example, the proposed condition for scalar
nonlocal bifurcations and numerics show complete and global agreement. 
We note that no deviation of the sort in Fig. \ref{Fig:mix} of the
first example occurs here because scalar nonlocal bifurcation boundaries 
do not intersect with local bifurcation boundaries of $v$-pulses now. 
In fact, no local bifurcations of daughter $u$-solutions from pure $v$-pulses
occur for $s<1$. 
This is because, unlike in the CNLS example
where the transformation $\omega \to 1/\omega$ mapped back curves into
the domain of interest ($0<\beta<\infty$), here this transformation
takes curves into the region $s>1$. 

\section{Linear instability of solitary waves arising from nonlocal bifurcations}
An important question about solitary waves generated by scalar nonlocal bifurcations
is their linear stability. These waves are 
multi-humped by construction. The common wisdom is that multihump
solitary waves in conservative systems are linearly unstable
(for instance, see \cite{Ya:01}). However, this is not always the case, 
as {\em stable} multihump solitary waves in the saturable coupled NLS
model (\ref{elenaPDE1}), (\ref{elenaPDE2}) have been reported
\cite{Osetal:99}. A comprehensive stability analysis of these solitary waves
is quite involved and lies outside the scope of the present article. 
However, in this section, we will selectively test the linear stability of 
a few such solitary waves as discussed above. These results are suggestive of
the linear stability behavior of solitary waves born out of scalar nonlocal bifurcations as a whole.   
 
Our strategy for testing the linear stability is by numerically simulating
the linearized equations of the two models (\ref{CNLS_PDE1}, \ref{CNLS_PDE2}) 
and (\ref{elenaPDE1}, \ref{elenaPDE2}) expanded around solitary waves.
For both models, we perturb solitary waves as
\begin{equation}
U(x, t)=e^{it}\left\{u(x)+\epsilon \psi(x,t)\right\}, \hspace{0.5cm} 
V(x, t)=e^{i\omega^2 t}\left\{v(x)+\epsilon \phi(x,t)\right\}, 
\end{equation}
where $(u, v)$ is a solitary wave, 
$\epsilon\ll 1$, and $(\psi, \phi)$ are perturbation functions. 
Substituting this perturbed solution into each model and dropping terms which are
order $\epsilon^2$ and higher, the linearized equations for $\psi$ and $\phi$ will
be obtained. Then we simulate this linearized system for a long time, starting with a random noise. 
If the solutions $\psi$ and $\phi$ exponentially grow, then the solitary wave $(u, v)$
is linearly unstable. 

\begin{figure}[p]
\begin{center}
\parbox{16cm}{\postscript{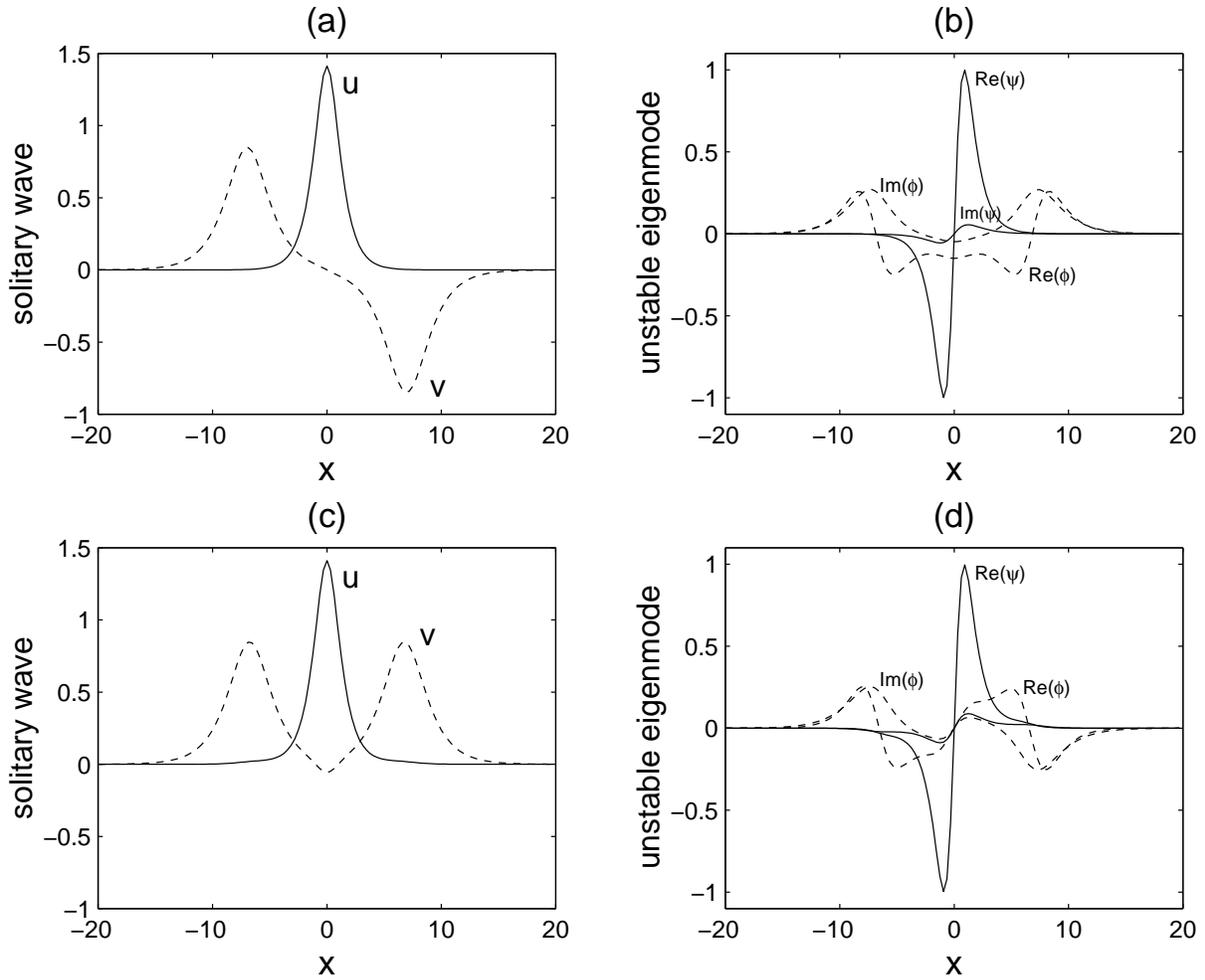}{1.0}}
\caption{Linear instability of two solitary waves born out of scalar 
nonlocal bifurcations in the system (\ref{CNLS_PDE1}, \ref{CNLS_PDE2}). 
(a, b) the solitary wave and its unstable eigenmode at $(\beta, \omega)=(0.29,0.6)$; 
(c, d) the solitary wave and its unstable eigenmode at $(\beta, \omega)=(1.6, 0.6)$.
} 
\label{Fig:new1}
\end{center}
\end{figure}

First, we consider the coupled NLS model (\ref{CNLS_PDE1}, \ref{CNLS_PDE2}). 
Its nonlocal bifurcation boundary is shown in Fig. \ref{Fig:fig1}. 
At $(\beta, \omega)=(0.29, 0.6)$ which is close to the nonlocal bifurcation
boundary with $n=1$, the solitary wave born out of this nonlocal bifurcation
is displayed in Fig. \ref{Fig:new1}(a) [see also Fig. \ref{Fig:safig}]. 
Numerical simulation of the linearized equation around this solitary wave
shows that there is a purely exponentially-growing unstable eigenmode. The 
unstable eigenfunction is displayed in Fig. \ref{Fig:new1}(b). 
The unstable eigenvalue is 0.083 (purely real). Thus, 
this solitary wave is linearly unstable. 
At $(\beta, \omega)=(1.6, 0.6)$ which is close to the nonlocal bifurcation
boundary with $n=2$ (see Fig. \ref{Fig:fig1}), another solitary wave
is born and displayed in Fig. \ref{Fig:new1}(c). 
We have found that this wave is also linearly unstable. The unstable eigenfunction
is shown in Fig. \ref{Fig:new1}(d), and the unstable eigenvalue is 0.13 (purely real). 

\begin{figure}[p]
\begin{center}
\parbox{16cm}{\postscript{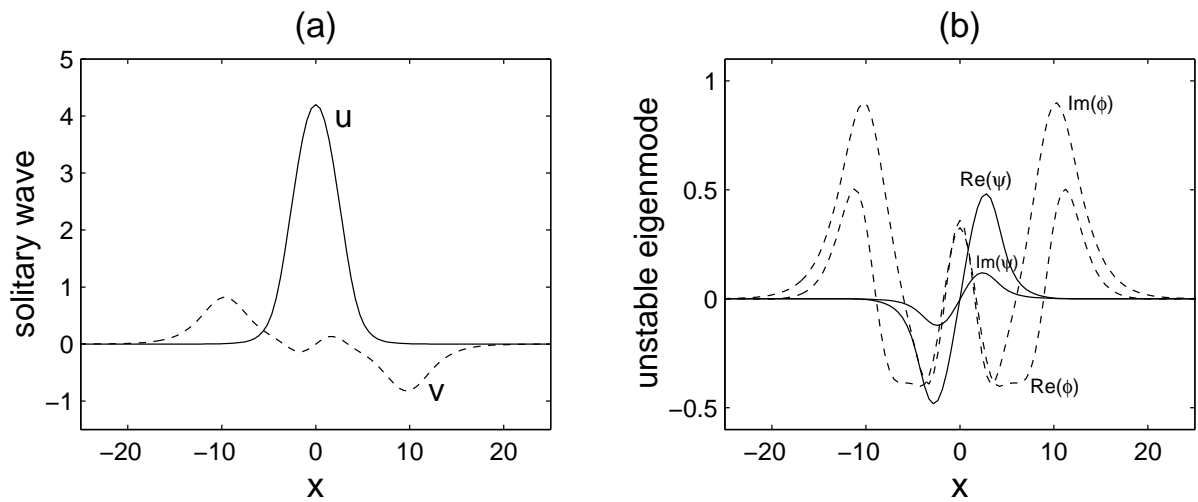}{1.0}}
\caption{Linear instability of a solitary waves born out of scalar 
nonlocal bifurcations in the system (\ref{elenaPDE1}, \ref{elenaPDE2}). 
(a) the solitary wave at $(s, \omega)=(0.81, 0.5)$; (b) the unstable eigenmode.
} 
\label{Fig:new2}
\end{center}
\end{figure}

Next, we consider the saturable model (\ref{elenaPDE1}, \ref{elenaPDE2}). 
As shown in Fig. \ref{Fig:Elenasep}, at $(s, \omega)=(0.81, 0.5)$, 
there is a solitary wave born out of a nonlocal bifurcation. This solitary
wave is reproduced in Fig. \ref{Fig:new2}(a). 
By simulating the linearized system, we have found that this solitary wave
is linearly unstable as well. The unstable eigenfunction is shown in 
Fig. \ref{Fig:new2}(b), and the unstable eigenvalue is 0.090 (purely real). 

From these selective numerical testings, we have reason to believe that
solitary waves which are just born out of scalar nonlocal bifurcations in 
the two models (\ref{CNLS_PDE1}, \ref{CNLS_PDE2}) 
and (\ref{elenaPDE1}, \ref{elenaPDE2}) are linearly unstable in general. 
This belief is consistent with a previously established fact that
solitary waves born out of {\em vector} nonlocal bifurcations are linearly unstable
\cite{Ya:01}. 
However, when these solitary waves have moved far away from the boundaries
of nonlocal bifurcations, instability may disappear, and the solitary waves
may become stable. In fact, we have found that in the saturable model
(\ref{elenaPDE1}, \ref{elenaPDE2}), when solitary waves of Fig. \ref{Fig:Elenasep}(a)
move to the local bifurcation boundary $s=s^{LB}$, they indeed become linearly stable
(a similar phenomenon has been reported in \cite{Osetal:99}). 
But in the CNLS model (\ref{CNLS_PDE1}, \ref{CNLS_PDE2}), solitary waves remain
unstable when they move to the boundaries of local bifurcations 
\cite{PeliYa00,Peli02}. 

\section{Discussion}
\label{discussion}

In this paper we have described by numerical computation, plausible
argument and detailed asymptotic analysis a new kind of bifurcation of
solitary waves for coupled nonlinear Schr\"{o}dinger systems. 
Our geometric argument suggests that this so-called scalar nonlocal
bifurcation occurs when the linearization of the nonlinear system
around a scalar pulse has purely growing asymptotics at infinity. 
Our matched asymptotic analysis confirms this condition. It further
reveals that when this scalar nonlocal bifurcation boundary 
intersects with certain local-bifurcation boundaries, 
the actual nonlocal bifurcation can turn from scalar to non-scalar 
at the intersection. All these theoretical results are fully supported
by our numerics on two coupled NLS systems (\ref{u}), (\ref{v}) and
(\ref{elena1}), (\ref{elena2}). 
Our direct numerical simulations suggest that 
solitary waves which are just born from scalar nonlocal bifurcations
are linearly unstable, but they can lead to stable solitary waves by
parameter continuation. 

It remains an open problem to fully unfold
the codimension-two nonlocal bifurcation that occurs when the scalar nonlocal
bifurcation boundary intersects the local-bifurcation boundary.
In particular it remains unknown whether a scalar non-local bifurcation
continues to occur on the upper portion of the solid
curve in Fig.~\ref{mix:copy} (upper left panel). 
We have found that the wave that starts from the
local bifurcation with $n=3$ does not end there, but this does not
exclude the possibility of other branches of vector solitons terminating
on this possible nonlocal bifurcation boundary. Preliminary numerical searching
has not revealed any candidate branches, but a more careful study is 
required.

Finally we should mention that the ideas in this paper were originally
motivated by the observation of what was termed a `jump' bifurcation
in \cite{KoBuStChSa:00}. There, a nonlocal bifurcation occurs where a
pure-$v$ pulse bifurcates at infinity from a vector soliton. The model
in question is for a third-harmonic generation system where the
symmetry is such that pure $u$ solutions do not exist. At present it
is not clear whether a simple criterion like the one in this paper can
be applied. One could conjecture that the jump
bifurcation occurs when linearisation around the central vector
soliton has precisely the growing exponential asymptotics in its tail
of the pure-$v$ pulse. But this conjecture requires careful 
numerical and analytical confirmations. 

\section*{Acknowledgments}
The authors acknowledge insightful conversations with Alice Yew
and Boris Malomed. 
This work was initiated while JY was visiting the UK under
an EPSRC grant. ARC also holds an Advanced Research Fellowship from
the EPSRC. JY's work was also supported by the US National Science Foundation
grant DMS-9971712 and The Air Force Office of Scientific Research contract
USAF F49620-99-1-0174.

\bibliography{reference}

\begin{thebibliography}{10}

\bibitem{Ak:95}
N.~N. Akhmediev, A.~V. Buryak, J.~M. Soto-Crespo, and D.~R. Andersen.
\newblock Phase-locked stationary soliton states in birefringent nonlinear
  optical fibers.
\newblock {\em J. Opt. Soc. Am. B}, 12:434, 1995.

\bibitem{AkAn:93}
N.N. Akhmediev and A.~Ankiewicz.
\newblock Novel soliton states and bifurcation phenomena in nonlinear fiber
  couplers.
\newblock {\em Physical Review Letters}, 70:2395--2398, 1993.

\bibitem{De:76b}
R.L. Devaney.
\newblock Reversible diffeomorphisms and flows.
\newblock {\em Trans. Amer. Math. Soc.}, 218:89--113, 1976.

\bibitem{Eleonsky}
V.M. Eleonsky, V.G. Korolev, N.E. Kulagin, and L.P. {Shil'nikov}.
\newblock Bifurcations of the trajectories at the saddle level in {H}amiltonian
  systems generated by two coupled {Schr\"{o}dinger} equations.
\newblock {\em Chaos}, 2:571--579, 1992.

\bibitem{HaSh:94}
M.~Haelterman and A.P. Sheppard.
\newblock Bifurcation phenomena and multiple soliton bound states in isotropic
  {Kerr} media.
\newblock {\em Phys. Rev. E}, 49:3376--3381, 1994.

\bibitem{KoBuStChSa:00}
K.~Kolossovski, A.V. Buryak, V.V Steblina, A.R. Champneys, and R.A. Sammut.
\newblock Higher-order nonlinear modes and bifurcation phenomena due to
  degenerate parameteric four-wave mixing.
\newblock {\em Phys. Rev. E}, 62:4309--4317, 2000.

\bibitem{LaLi:77}
L.D. Landau and E.M. Lifshitz.
\newblock {\em Quantum Mechanics: non-relativistic thoery}.
\newblock Pergamon Press, Oxford, 3rd edition, 1977.

\bibitem{Ma:73}
S.V. Manakov.
\newblock On the theory of two-dimensional stationary self-focusing of
  electromagnetic waves.
\newblock {\em Zh. Eksp. Teor. Fiz.}, 65:1392, 1973.

\bibitem{MiSeCh:98}
M.~Mitchell, M.~Segev, and D.N. Christodoulides.
\newblock Observation of multihump multimode solitons.
\newblock {\em Phys. Rev. Lett.}, 80:4657, 1998.

\bibitem{OsKi:99}
E.A. Ostrovskaya and {Yu.S.} Kivshar.
\newblock Multi-hump optical solitons in a saturable medium.
\newblock {\em J. Opt. B: Quantum Semiclass. Opt.}, 1:77--83, 1999.

\bibitem{Osetal:99}
E.A. Ostrovskaya, {Yu.S.} Kivshar, D.V. Skryabin, and W.J Firth.
\newblock Stability of multi-hump optical solitons.
\newblock {\em Phys. Rev. Lett.}, 83:296--299, 1999.

\bibitem{Peli02}
D.E. Pelinovsky.
\newblock Matrix stability theory for incoherent optical solitons.
\newblock {\em Preprint}.

\bibitem{PeliYa00}
D.E. Pelinovsky and J.~Yang.
\newblock Internal oscillations and radiation damping of vector solitons.
\newblock {\em Stud. Appl. Math.}, 105:245, 2000.

\bibitem{Ya:97}
J.~Yang.
\newblock Classification of the solitary wave in coupled nonlinear
  {Schr\"{o}dinger} equations.
\newblock {\em Physica D}, 108:92--112, 1997.

\bibitem{Ya:98}
J.~Yang.
\newblock Multiple permanent-wave trains in nonlinear systems.
\newblock {\em Stud. Appl. Math.}, 100:127, 1998.

\bibitem{Ya:01}
J.~Yang.
\newblock Interactions of vector solitons.
\newblock {\em Phys. Rev. E}, 64:026607, 2001.

\bibitem{Be:96}
J.~Yang and D.J. Benney.
\newblock Some properties of nonlinear wave systems.
\newblock {\em Stud. Appl. Math.}, 96:111--139, 1996.

\bibitem{Ye:01}
A.C. Yew.
\newblock Multipulses of nonlinearly coupled schrodinger equations.
\newblock {\em J. Diff. Eqs.}, 173:92--137, 2001.

\bibitem{YeChMc:99}
A.C. Yew, A.R. Champneys, and P.J. McKenna.
\newblock Localised solutions of a coupled first and second harmonic nonlinear
  shr\"{o}dinger system.
\newblock {\em J.\ Nonlin.\ Sci.}, 9, 1999.

\bibitem{YeSaJo:00}
A.C. Yew, B.~Sandstede, and Jones C.K.R.T.
\newblock Instability of multiple pulses in coupled nonlinear schrodinger
  equations.
\newblock {\em Phys. Rev. E}, 61:5886--5892, 2000.

\bibitem{Za:71}
V.E. Zakharov and A.B. Shabat.
\newblock Exact theory of two-dimensional self-focusing and one-dimensional
  self-modulation of waves in nonlinear media.
\newblock {\em Zh. Eksp. Teor. Fiz.}, 61:118, 1971.

\end{thebibliography}

\end{document}